\documentclass{aastex}
\shorttitle{Merging of globular clusters in galactic nuclei}
\shortauthors{Capuzzo-Dolcetta \& Miocchi}

\def\beq{\begin{equation}}
\def\eeq{\end{equation}}
\def\schw{Schwarzschild's }

\def\media#1{\langle #1\rangle}
\def\a{(a1)}
\def\b{(b1)}
\def\C{(c1)}
\def\d{(d1)}
\def\al{(a2)}
\def\bl{(b2)}
\def\cl{(c2)}
\def\dl{(d2)}

\def\msol{\mathrm{\ M}_\odot}
\def\media#1{\langle #1\rangle}
\begin{document}

\title{Merging of globular clusters within inner
galactic regions.\\
II. The Nuclear Star Cluster formation
}

\author{R. Capuzzo-Dolcetta}
\email{roberto.capuzzodolcetta@uniroma1.it}
\author{P. Miocchi}
\email{miocchi@uniroma1.it}
\affil{Dipartimento di Fisica, Universit\'a di Roma ``La Sapienza",\\
P.le Aldo Moro, 2, I00185 -- Rome, Italy.}

\begin{abstract}
In this paper  we present the results of two detailed
N-body simulations of the interaction of a sample of four massive globular
clusters in the inner region of a triaxial galaxy. A full merging of the clusters takes
place, leading to a slowly evolving cluster which is quite similar to observed Nuclear Clusters.
Actually, both the density and the velocity dispersion profiles match qualitatively,
and quantitatively after scaling, with observed features of many nucleated galaxies.
In the case of dense initial clusters, the merger remnant shows a density profile more
concentrated than that of the progenitors, with a central density 
higher than the sum of the central progenitors central densities.
These findings support the idea that a massive Nuclear Cluster may have formed in 
early phases of the mother galaxy evolution and lead to 
to the formation of a nucleus, which, in many  galaxies, has 
indeed a luminosity profile similar to that of an extended King model.
A correlation with galactic nuclear activity is suggested.
\end{abstract}


\defcitealias{miocchi05}{Paper I}

\section{Introduction}
In the first paper of this series devoted to the study of the interaction of
globular clusters (GCs) in triaxial galaxies, we analyzed the face-on
collision between two GCs moving on quasi-radial orbits in the central galactic
region, with the aim, also, to understand how effective is the
tidal distortion.
Our first finding was that the face-on collision is practically uneffective
respect to the role of the external tidal field. Actually, the tidal erosion
has been shown to be such to destroy {\it loose} GCs (initial King
concentration parameter $c<1$) in few passages through the galactic centre,
while \textit{tight} cluster ($c\geq 1.6$) keeps bound a substantial amount of their
mass up to the complete orbital decaying. At this latter regard, another
important result is that the orbital energy dissipation due to the tidal
interaction is of the same order of that caused by dynamical friction.
At the light of these results, and given that dynamical friction was shown to
be important in segregating massive GCs in triaxial potentials
(\citealt*{pesce}; \citealt{capdol93}; \citealt{cv05}), a natural further step in the present program of
investigation is to study the possible merging of a set of GCs decayed in the
central galactic region to see whether a sort of \lq super star cluster \rq
(SSC) results from the merging, and what its morphological and dynamical
characteristics could be.

It is quite ascertained the existence of very bright ($10^7\div 10^8 $L$_\odot$)
clusters of stars in the central region of galaxies across the Hubble
sequence thanks to VLT and HST observations \citep{walch05,vandermarel,wehner}.
On the basis of integrated colours and of the estimated $M/L$ ratios, these
clusters are usually thought to be young or, at least, to contain a
significant population of young stars. In any case, that of the age is a
controversal point, because, for instance, the nuclear star clusters (NCs) in M82
show evidence for mass segregation despite their spectra are well fitted by
stellar population synthesis models with ages $10$--$50$ Myr \citep{McCr}.
\citet{bok04} fitted analytical models to Hubble Space Telescope images
of 39 NCs in order to determine their effective radii after
correction for the instrumental point-spread function.
They compared the luminosities and sizes of NCs to those of
other ellipsoidal stellar systems, in particular the Milky Way globular
clusters, finding for NCs a narrow size distribution statistically
indistinguishable from that of
Galactic GCs, even though the NCs are, on
average, 4 mag brighter than the old GCs. They discuss some
possible interpretations of the similarity among NCs and Galactic GCs and,
from a comparison of NCs luminosities with various properties of
their host galaxies, find that more luminous galaxies harbor more luminous
NCs.
It remains unclear whether this correlation reflects the influence of
galaxy size, mass, and/or star formation rate.

However, \citet{rossa06}, by means of spectroscopic HST/STIS data,
derived NCs ages for 19 galaxies and found that they
range from 10 Myr to 10 Gyr, with a non negligible
presence of old clusters.
A comparable result has been deduced also for 9 NCs of very late-type,
bulge-less spirals, by \citet{walch06} by means of a high-resolution
spectroscopic survey with VLT/UVES.

\citet{genz03}, in their study of the stellar cuspy distribution around the
Supermassive Black Hole (SBH) in the Galactic center, found that the K-band
luminosity function of the local NC (within 9'' of Sgr A*) resembles that of
the large-scale Galactic bulge except for showing an
excess of stars at $K_s \leq 14$. It fits with population synthesis models of an old,
metal-rich stellar population with a contribution from young, early, and
late-type stars at the bright end.
In the central arcsecond, they argued that a stellar merger model is the
most appealing explanation. These stars may thus be ``super blue-stragglers'',
formed and ``rejuvenated'' through mergers of lower mass stars
in the very dense  ($\geq 10^8 \msol$ pc$^{-3}$) environment of the cusp.

Another intriguing piece of the puzzle of the structure of SSCs is given by
\citet{baumg03}. Through a comparison between the observational data on
the kinematical structure of the very bright cluster G1 in M31 ---obtained 
with the HST WFP Camera 2 and Space Telescope Imaging
Spectrograph instruments--- and their results of dynamical simulations 
carried out using the special purpose computer GRAPE-6, they obtained very good 
fits when starting simulations with initial conditions extracted from
the end product of a previous simulation concerning a merging between
two pre-existent star clusters.
The merging explains observed features without the need to invoke the 
presence of an intermediate-mass black hole in the center of G1.

Not many simulations have been presented in the literature that study the
possible formation scenarios for SSCs. Among these, we remind those by 
\citet{fell02,fell05} finding that star clusters aggregates, like the ones found
in the Antennae or Stephan Quintet, are very likely the merger progenitors
of SSCs. 
Interestingly enough, these authors also find that the resulting SSC is a stable
and bound object, whose density profile is well fitted by a King profile, even if the
mass loss of the merger product occurs through every perigalacticon passage.

\section{Globular cluster merging and galactic nuclei}
The increasing amount of data about massive clusters in 
galaxies, together with that of GC systems in galaxies, especially of
early type but
also for spirals \citep[][etc.]{harris86,harris91,ashzepf98,harris01} makes
interesting the investigation, started in the first paper of this series
\citep[][hereafter \citetalias{miocchi05}]{miocchi05}, of the fate of
massive GCs moving in the parent galaxy field, subjected to dynamical friction
braking and tidal interaction.
In \citetalias{miocchi05} we studied the dynamical evolution of two GCs with
mass $\approx 10^7$ M$_\odot$ moving on quasi radial orbits in a triaxial 
galactic potential, following eight passages across the galactic center. The scope of 
that paper was to investigate the chances of survival of GCs to possible tidal disruption 
induced both by the external field and by mutual cluster-cluster interaction; with this 
aim, we maximized these effects by giving the clusters initial conditions
corresponding to quasi-radial orbits.

In this Paper II, we have another aim, that is to study whether and how 
the merging of various massive GCs decayed by dynamical friction in the inner galactic 
region may occur.
The main questions to answer are:
(i) given some (realistic) initial conditions for a set of
GCs which experienced a significant orbital decay, are they undergoing a 
full merge?
(ii) if so, what is the time needed? (iii) what is the final structure 
of the merged NC?
(iv) does it attain a quasi-steady state?

The answers to these questions are of overwhelming importance to give 
substance to the interpretation of the formation of early type galaxy nuclei via
merging of decayed GCs, a hypothesis raised first by \citet{tremetal75} and
subsequently extensively studied by \citet{capdol93,ct97,ct99,cv05}. 

This scenario of galactic nuclei formation has been raised again
recently, due to the independent and almost contemporaneous findings
that NC masses obey to similar scaling relationships with host galaxy properties as
SBHs do (see \citealt{rossa06} for spirals; \citealt{wehner} for dwarf elliptical [dE]
galaxies and \citealt{cote06} for elliptical galaxies).
In particular, \citet{cote06} give evidence, in early-type galaxies in the Virgo cluster,
of NC luminosity distributions that are much
better fitted by an extended profile (King's) rather than by a point source
(see \citealt{bok02} for a similar finding in late-type
spirals and \citealt{guzman} for dE galaxies in the Coma Cluster).
The half mass radii of nuclei ($r_h$) are in the range  $2 < r_h$(pc) $ < 62$, 
with $\media{r_h} = 4.2$ pc and correlate with the nucleus
luminosity: $r_h \propto L_n^{0.5\pm 0.03}$.
The mean of the frequency function for the nucleus-to-galaxy luminosity ratio in
nucleated galaxies, $\log\eta = - 2.49 \pm 0.09$ is indistinguishable from
that of the Super Massive BH-to-bulge mass ratio,
 $\log (M_{BH}/M_{gal}) = -2.61 \pm 0.07$, calculated in 23 early-type galaxies. 
On these bases, \citet{cote06} argue that resolved stellar nuclei are 
the low-mass counterparts of nuclei hosting SBHs  detected in the bright
galaxies. 
If this view is correct, then one should think in terms of central massive
objects, either SBHs or SSCs, that accompany the 
formation and/or early evolution of almost all early-type galaxies.

It is clear that these characteristics of galactic nuclei well fit
into a scenario of multiple GC merging in the inner galactic regions (that we
call ``dissipationless'' scenario) as alternative to the ``dissipational''
scenario, this latter being based mainly on speculative hypotheses
\citep[see, e.g.,][]{vdb86,silkws87,babrees92} supported by some quantitative
results \citep{mihhernq94}.
The proof of the validity of the ``dissipationless'' hypothesis requires,
at first, a detailed
$N$-body simulation of the interaction of stellar clusters in the inner region of
their parent galaxy, taking into account both the cluster-cluster and the
cluster-galaxy dynamical interaction. This latter includes tidal distortion,
acting on the cluster internal motion, and dynamical friction, acting on the 
cluster orbital motion.

In this context, we cite the encouraging
results of the numerical simulations made by \citet{oh00} who revisited the
hypothesis of dE nuclei formation through the orbital decay of GCs
and suggested that this occurs mainly in galaxies with a relatively
weak extragalactic tidal perturbation, leading to the formation of compact nuclei
within a Hubble timescale. They also show that the central galactic field does not
destroy the integrity of the clusters and facilitates the merging to occur.
Moreover, they find that the observed central structures of
some nucleated Virgo Cluster dEs are well reproduced by superimposing
a small number of globular clusters to the galactic stellar distribution.
Nevertheless, \citeauthor{oh00} used a very small number of particles ($=500$ in
each cluster) to simulate the final merger stage, during which they
neglected the role of dinamical friction, as well. Thus, their results could be
affected by spurious collisional effects, even though a rather large smoothing
radius was adopted in the inter-particle interactions.

More recently, \citet{bekki04} show how the formation of NCs via multiple merging of 
GCs, leads to systems with global scaling relations which are
compatible with those observed for galactic nuclei.
Unfortunately, as mentioned before, these results cannot be considered conclusive
because in these authors' simulations the, important, role of the external field
was neglected.

\section{The simulations}
\label{model}
We consider GCs as $N$-body systems moving within a triaxial
galaxy represented by an analytical potential,
subjected also to the deceleration due to dynamical friction (hereafter df).
To keep a high level of resolution in the simulations (i.e., a large number of
particles to represent each GC without exceeding computational capabilities,
we decided to limit our study to the interaction of a limited number, four,
of GCs.

We studied the merging process of these GCs, as if they were already decayed
to the inner region of the galaxy, in two separate simulations (cases 1 and 2).
A quick orbital decay induced by df is due to initial large values for the
total mass of the clusters. These large masses
are compatible with those of many young GCs actually observed in various
galaxies \citep[e.g.,][and references therein]{kissler}, see \citetalias[][for
a deeper discussion]{miocchi05}.
The 4 clusters are labeled \a, \b, \C, \d\ in the case 1 and
\al, \bl, \cl, \dl\ in the case 2.
In the simulation of case 2, the 4 GCs have the same
initial orbital conditions but their scaling parameters are such to give
them less dense and more extended initial configurations (see also Table \ref{tab1}
and Sect.\ref{clumod}).

Unless otherwise specified, the same units of measure used in
\citetalias{miocchi05} are adopted in this paper: lengths, masses and time
are measured, respectively, in unit of the galactic core radius $r_b$, of the
galactic core mass $M_b$ and of the galactic core-crossing time
$t_b\equiv [r_b^3/(GM_b)]^{1/2}$.
Note that the actual physical values for these quantities are unrelevant because
the results of simulations can be scaled as long as the 2-body stellar collisions
are negligible; otherwise,
the dynamics would obviously depend also on the mass of the \emph{\/individual} star
in the clusters.



As regards the computational techniques, we adopted an $N$-body
representation for the stars in the clusters, simulating their
dynamics by means of the parallel `treeATD' code \citep{bib2}, whose main features
were also described in \citetalias{miocchi05}.


\subsection{The galactic model}
\label{bulgemodel}
We adopt the same galactic model used in the simulations of
\citetalias{miocchi05}, i.e. the self-consistent triaxial
model described in \citet{zeeuw}.
It corresponds to a non-rotating ellipsoidal and triaxial distribution of matter
oriented in such a way to have the longest and the shortest principal axis
aligned with the coordinate axes $x$ and $z$,
respectively.
The axial ratios are $2$ : $1.25$ : $1$, leading to
a projected profile in agreement with that observed in the spheroids
of some spirals \citep[see, e.g.,][]{bertola91,matthews04} and in
elliptical galaxies \citep[see, e.g.,][]{wagner88,davies01,statler04}.
The resulting potential can be expressed as the sum of a spherically simmetric
density following the modified Hubble's law
\beq
\rho_b(r)=\rho_{b0} \left[1+\left(\frac{r}{r_b}\right)^2\right]^{-3/2},
\label{hubble}
\eeq
with scale parameters $r_b$ (core radius) and $\rho_{b0}=M_b/r_b^3$, plus
other two
non-spherical terms that give the triaxial behaviour \citep{schw}.
The mass parameter $M_b$ is the mass enclosed
in a radius slightly smaller than $r_b$, in fact
$M_b\simeq 0.45M(r_b)$.
The mass in the generic sphere of radius $r$ is contributed only by the
spherically simmetric part, $\rho_b(r)$, of total density, giving:
\begin{displaymath}
M(r)=4\pi M_b\left\{ \ln \left[ \frac{r}{r_b}+\left(\frac{r^2}{r_b^2}+1
\right)^{1/2}\right] \right.
\end{displaymath}
\beq
\ - \left.\frac{r}{r_b}\left(\frac{r^2}{r_b^2}+1\right)^{-1/2}\right\},
\label{m_di_r}
\eeq
that gives an infinite total mass.  However, there is no need of a cutoff
in the model, because
only the gravitational potential (which is not divergent) is important for the
purpose of our simulations.

As in \citetalias{miocchi05}, the reference frame has the origin at the
galactic center and the $x$ and $z$ axes are, respectively, along the
maximum and minimum axis of the triaxial ellipsoid.
The centers-of-mass (CM) of the clusters were initially located well
within the galactic core (see Table~\ref{initialorbits}).

With regard to the df, we used the generalization  to the triaxial case
\citep[see][]{pesce}, of the Chandrasekhar formula \citep{chandra},
with a self-consistent evaluation of the velocity dispersion
tensor, taking also into account that the GCs are
extended objects. See \citetalias{miocchi05} for more details
at this regard.

\subsection{The cluster model}
\label{clumod}
Our numerical study involved two sets of four clusters evolved in two separate
simulations. The GCs initial internal
distribution was sampled from a stellar \citet{king66}
isotropic model with total mass $M$, velocity parameter\footnote
{Which is such that, at the center, $\sigma^2\equiv\langle v^2\rangle\simeq 3\sigma_\mathrm{K}^2$, the
approximation being as better as the model concentration is higher
\citep[][sect. 4.4c]{bib4}.}
$\sigma_\mathrm{K}$, limiting and ``King'' radius $r_t$ and $r_c$, respectively;
$c=\log (r_t/r_c)$ is the concentration
parameter, $t_{ch}= [r_h^3/(GM)]^{1/2}$ the half-mass crossing time
and $\rho_0$ is the central density.
The `limiting radius' is the radius at which the King
distribution function drops to zero, to reproduce the presence of the
external field \citep{king66}.
The initial values of these parameters are listed in Table \ref{tab1}.

Each cluster was represented with $N=2.5\times 10^5$ `particles',
whose individual masses were assigned according to a Salpeter's
mass distribution ($dN\propto m^{-2.35}dm$) cutted at $m_{\min}$
and $m_{\max}=100 m_{\min}$, with average $\langle m\rangle =3.1m_{\min}$,
where $m_{\min}$ is adjusted to give the desired cluster total mass (see Table
\ref{tab1}). When simulating massive GCs, the assumption $N=2.5\times 10^5$ gives a high
value of $m_{\min}$ that shifts the chosen mass distribution 
in form of Salpeter law towards large masses so to lose its representativity
for a real GC stellar mass spectrum.
However, as discussed in \citetalias{miocchi05}, we use unequal mass particles
to verify that no collisional relaxation
takes place during the simulation, as demanded by the 2-body relaxation time
of the clusters which is in any case longer than the time to merge.

Actually, even for the most compact case considered here
(model \a\ see Table \ref{tab1}), the half-mass relaxation time of the simulated cluster
is larger than the merging time $t_m\sim 18$ (see Sect. \ref{results}) after which the 
clusters lose their individuality
and evolve as a larger $N$ system (see discussion in Sect.\ref{stationary})
The real cluster (with a larger number of stars) relaxation time is
obviously longer than that of the simulated (sampled) cluster.

Were the GCs massive enough to decay rapidly into the galactic core,
it can be safely assumed (see also \citetalias{miocchi05}, sect. 2)
that their age at the beginning of the 
simulation is less than their internal 2-body relaxation time, so 
that mass segregation has not occurred and, thus, it was not
included in the initial stellar distribution. 

In the following, we often refer to the system \emph{\/center-of-density}
(CD) -- i.e. the average particle position weighted with the local density
instead of the mass --
as defined in \citet{cashut}.
As in \citetalias{miocchi05}, in most cases we took the CD
as the origin of the best suitable reference frame for the study of the
internal system properties.


\section{Results}
\label{results}
 
In spite that the true conserved quantity, in absence of any external dissipation,
is the center-of-mass orbital energy, it is preferable to refer to the CD orbital energy
instead, because the CD identifies better the actual GC location when
the outer part of the cluster is highly distorted and dispersed.
Defining $E_{orb}$ as the orbital energy (per unit mass) of the
cluster CD, its dissipation due to both the df braking effect and to the
tidal interaction with the environment, can be quantified
by the following parameter
\citepalias{miocchi05}:
\begin{equation}
\xi_{orb}(t)=\frac{E_{orb}(t)-\Psi_0}{E_{orb}(0)-\Psi_0},
\end{equation}
where $\Psi_0=-4 \pi GM_b/r_b$ is the central galactic potential well.
In Fig.~\ref{decadorb}, $\xi_{orb}(t)$ is plotted for all the clusters in the
two cases studied.

The merging occurs rather quickly, as it is indicated by the snapshots of the system
as projected onto one of the coordinates plane (see Fig.s~\ref{snap1} and \ref{snap2}
for the case 1) and clearly seen from the time
behaviour of the Lagrangian radii of the whole system (Fig.~\ref{rlag}). Lagrangian
radii attain steady values at a ``merging'' time of about $t_m\sim 18$, when also the
innermost radius shown stabilizes; note, that the inner is
the Lagrangian radius, the later is the time at which the stable state is
achieved. A similar merging time can be deduced also by the time behaviour of
the distances of the clusters CD to the galactic center (Fig.~\ref{distance}).

Of course, the energy dissipation rate is closely related to the
prescription adopted for the evaluation and treatment of df.
For instance, our evaluation \citepalias[][Appendix A]{miocchi05} could
lead, in principle, to an overestimate of the effect, because we consider the cluster
as a constant mass point\footnote{To partially overcome this problem we conservatively used a
halved cluster mass as mass
parameter in the df formula.}.
Obviously, df would be accounted for
accurately when adopting a full $N$-body representation
of the galactic region in which GCs move;
for this reason we are running a self-consistent simulation
that we will present in a forthcoming paper. However, preliminary results
show a clearly \textit{shorter} orbital decaying time for the
same clusters of case 1, but moving in a self-consistent
representation of the same galaxy sampled by $5\times 10^5$ ``particles''.
This is shown by the time behaviour of the distance
of the clusters CD from the galactic center depicted in Fig.\ref{distance_full},
from which a merging time $t_m\sim 13$ is deduced, i.e. a factor $1.4$ smaller than in
the simulations with the analytic df form.
This finding agrees with some recent results that show
how stars stripped from the cluster by the field, but still close enough
to the system, continue to contribute to mass of the decelerating system \citep{fell07}.
Moreover, recent fully self-consistent $N$-body simulations confirm that the real
df effect can be even stronger that that estimated by the usual Chandrasekhar formula
\citep[probably for the further friction due to tidal effects,
see][]{funato}.


\subsection{The NC morphology}
To study the NC morphology, we evaluated the eigenvalues $\{I_1, I_2, I_3\}$
and the eigenvectors $\{\mathbf{e}_1,\mathbf{e}_2,\mathbf{e}_3\}$ of
the inertia tensor
\begin{displaymath}
I_{ij}=\frac{1}{N}\sum_{k=1}^N(x^k_i-x^0_i)(x^k_j-x^0_j)-
\frac{1}{N^2}\times
\end{displaymath}
\beq
\ \sum_{k=1}^N(x^k_i-x^0_i)\sum_{k=1}^N(x^k_j-x^0_j),
\label{inertiatensor}
\eeq
with $(x^k_1,x^k_2,x^k_3)$ being the position vector of the $k$-th particle,
and $(x^0_1,x^0_2,x^0_3)$ the CD position. In Fig.~\ref{eigen}, the square root
of the eigenvalues are plotted as a function of time.
They are evaluated considering, in
Eq. (\ref{inertiatensor}), only particles closer than $r_h$ to the system CD,
where $r_h$ is the half-mass radius of the NC at the end of the simulations.
Since $\sqrt{I_i}$ is equal to the length of the $i$-th axes of the
ellipsoid fitted to the matter distribution, then their ratios
give a direct measure of the triaxiality of the system;
e.g. if $I_1=I_2=I_3$ then the system is spherical.

It can be seen that, after the initial
merging phase, the NC as a whole attains a stable configuration.
Only  in the case 2 (Fig.~\ref{eigen}, upper panel) the system shape is moderately
axisymmetric (oblate) and, by examining the orientation of the eigenvectors,
it was found to have principal axes nearly aligned to those
of the galactic potential, with axial ratios
$\sqrt{I_1}$ : $\sqrt{I_2}$ : $\sqrt{I_3}\simeq
1.4$ : $1.4$ : $1$,
and with $\mathbf{e}_3$ nearly parallel to the $z$-axis.
This alignment suggests that the
final morphology is influenced, to some extent, by the galactic morphology.
Thus, it is not surprising that in the NC$_2$ outskirts the stars
tend, eventually, to move on orbits compatible with the underlying
galactic triaxial configuration,
as we will see from the velocity anisotropy (Sect. \ref{veldistr}).
Nevertheless, the \emph{\/self-gravity} of the simulated system is
important enough to make the shape of the NC$_2$ closer to a spherical configuration
than the galaxy.
Accordingly, the configuration of the NC$_1$ is almost spherical due to its
stronger self-gravity (Fig.~\ref{eigen}, lower panel).

\subsubsection{Density profile}
The radial density profile of the NCs, which we have seen to be quasi-spherical
at the end of the merging, is plotted in Fig.~\ref{dens}, where, for comparison,
the profile corresponding to the spatial superposition of the 4 progenitor
clusters, in their initial configurations, is reported, too.
This gives an immediate `visual' indication of the efficiency of the
merging
process; the merging remnant density profile is even more concentrated than that
given by the mere sum of the initial profiles of the four progenitors.
Actually, the merger profile has a smaller core (showing a central density a
factor $1.25$ higher) and a more extended envelope.
In fact, it is remarkably well fitted by a high-concentration ($c=2.2$)
King profile (see Fig.~\ref{dens}).
This result --- which agrees with what found in the different context of
merging galaxies since the pioneering \citet{white78} simulations ---
suggests that a sort of violent relaxation took place during the merging.
This relaxation is likely due to the rapidly varying potential acting on each
star of the clusters. 

In the initial cluster models a stellar mass spectrum was included without mass
segregation, in order to investigate possible relaxation phenomena due to either
spurious collisional effects or induced by the external tidal field.
The analysis of the behaviour of the average mass as a function of the
distance from the center in the final NC configuration
indicates \emph{\/no} evidence of mass segregation, thus enforcing the above-mentioned
hypothesis about the violent relaxation as a cause of the increased
concentration of the merging product.

\subsubsection{Velocity distribution}
\label{veldistr}
To compare our simulation results with observational data giving the radial
profile of the line-of-sight velocity dispersion, we must evaluate the ``total''
velocity dispersion,
$\sigma_{tot}$, by properly accounting for the contribution of the underlying galaxy,
this latter being given directly by the self-consistent model (according to the fitting
formulae [A3] and [A4] in \citetalias{miocchi05}).
Fig.\ref{veldisp} shows clearly a behaviour of $\sigma_{tot}$ decreasing  towards the
galactic center, where the surface density of the NC dominates and thus
$\sigma_{tot}$ lowers to the velocity dispersion of the NC, which is `colder' than the
host galactic core.
The behaviour shown in Fig.\ref{veldisp} may seem peculiar for a self-gravitating system,
which normally keeps its equilibrium through a negative (or null, if isothermal) gradient
of $\sigma^2$. This apparent peculiarity is due to that the central ``observed'' velocity
dispersion is dominated by the NC which is not a self-gravitating system, indeed.
This feature was actually found by a
Keck II spectroscopic analysis \citep[][see their Fig. 5]{geha}
for most of the Virgo cluster nucleated dEs of the sample.
The same $\sigma_{tot}$ radial behaviour is also shown by the solution of the Jeans equations
for a sample of NCs observed in late-type spirals \citep{walch05} and is also 
deduced on the basis
of direct $N$-body simulations by \citet{oh00} concerning GCs merging at the center
of dEs. This latter simulation confirms the occurrence of a merging event within
the central potential well of the host galaxy, even if care is due to the 
low number of particles that forces them to adopt a rather large gravitational
smoothing radius, in order to suppress fictitious collisional effects.  

As regards the velocity anisotropy, we studied the anisotropy parameters
$\beta_{\phi,\theta}=1-\sigma^2_{\phi,\theta}/\sigma^2_r$
of the stellar velocity distribution in the NCs, as a function of the
distance from the system CD, where $\phi$ and $\theta$ denote
the azimuthal or the polar angle, respectively.
These parameters are compared with those of the host galaxy.

We have found that  NC$_1$ has a nearly isotropic
velocity distribution ($|\beta_{\phi,\theta}|< 0.2$), while the influence of the
external galactic potential induces a $\beta_\phi<\beta_\theta <0$
(i.e. $\sigma_\phi^2 >\sigma_\theta^2 >\sigma_r^2$) in the outer region
 ($r\gtrsim 0.5r_b$) of the less compact NC$_2$, which means that the velocity
distribution in the outskirts is preferentially
tangential and parallel to the $xy$ plane, as in
the galactic model.
Noticeably, \citet{geha} found that the $\sigma_{tot}$ decrease in the
center of their observed nucleated dE can be reproduced, in most cases, by a model having
a nearly isotropic or moderately tangential velocity distribution.

Another relevant parameter is the central value of $\rho/\sigma^3$, which is
proportional to the density in the phase-space. Its value (see Table \ref{tab2}) in the
merged systems is smaller than that of the four individual progenitor clusters; this corresponds
to the, expected, rarefaction in the phase-space of the inner part of the system after the
mutual interaction of the merging clusters and, mainly, that with the external field.
This time-dependent interaction leads to a certain diffusion in phase space even if the
four GCs constitute a collisionless systems.
Note how the rise of the central space
density respect to the linear summation of individual GC densities in the case of
NC$_2$ is strongly balanced by expansion in velocity space
($\sigma^3$ of NC$_2$ is $\sim 5$ times higher than the average velocity dispersion
of the progenitors) so to decrease for a factor
$0.4$ the phase-space density.

As regards the virial equilibrium of NCs, it is worth noting that the King model
fitting of
NC$_1$ (Fig.\ref{dens}) gives a velocity parameter
$\sigma_\mathrm{K} = (4\pi G\rho_0)^{1/2}r_c/3\simeq 0.36$, significantly lower than
the true central value of the line-of-sight velocity dispersion
$\sim 0.52$.
This discrepancy is due to the non negligible influence of the external galactic potential,
which kinematically ``heats up'' the NC.
This has an important implication on the reliability of the NC mass determination by means
of the virial relation $M=\eta \sigma_p^2r_h/G$, with $\sigma_p$ the projected
 velocity dispersion. The true mass value for our NCs
is obtained with a virial parameter $\eta\simeq 2$ that is about 5 times lower than the
usually adopted values based on the unrealistic assumption of isolated King models.
This finding is in remarkable agreement with the results of \citet{walch05}
on the dynamical mass estimate of NCs observed
in bulge-less spiral galaxies. 

\section{Are resolved galactic nuclei compatible with 
a merging origin?}
An important result of this work is that a quasi-stationary NC may form as merger 
product of orbitally decayed massive clusters. The projected density profile in the
2 cases studied here is given in Fig.~\ref{dens}.
Its similarity to the resolved nuclear profiles of many of the galaxies 
observed by, e.g., \citet{bok02, geha, walch05, cote06} is clear;
of course, the ratio between the central
surface total density and that of the galaxy central plateau depends on the 
number of merged GCs. This ratio is $\approx 18$  for the four merging GCs of case 1
and $\approx 1.5$ for case 2.
An analysis of Fig. 4 in \citet{cote06} indicates ratios of the central luminosity
and the inward extrapolated galactic luminosity in the interval
$1 \leq L_0/L_{gal} \leq 100$. 
In the assumption that the density contrast scales linearly with the number of
GCs merged ($N_m$) the range of the observed ratio corresponds to the interval 
$0\leq N_m \leq 22$ for case 1 and $0\leq N_m \leq 267$ for the case 2. 
Of course, the linearity is not guaranteed; actually, the modes of the 
merging of a more abundant sample of GCs deserves a careful, forthcoming, study. 
Moreover, a fully self-consistent study of the feedback among the merging GCs and 
the surrounding stellar field is needed to confirm the modes and time of orbital decay 
and merging. This study will constitute a check of the reliability of the dynamical 
friction treatment done here in the frame of the \citet{pesce} scheme, which, if proved, 
will result on the possibility of saving a huge amount of CPU time in future simulations.

\section{The merger remnant stationariety}
\label{stationary}
In both the cases studied here the merging occurs at $\sim 20$ $t_b$ and the
resulting NC keeps its characteristics almost unchanged over the
remaining simulated time (other $25$ $t_b$). This does not imply, necessarily, the
secular stationariety of the merged system, which should be investigated 
extending sufficiently in time the $N$-body simulations.
Anyway, simple considerations on the relaxation times convince 
ourselves that the lifetime of the merged system should be quite long.
Actually, the half mass relaxation time of the cluster is \citep{spitzer}
\beq
t_{rh}=\frac{N}{7 \ln N}\left(\frac{M_b}{M}\right)^{1/2} 
\left(\frac{r_h}{r_b}\right)^{3/2} t_b
\eeq
which gives $t_{rh}=1.1\times 10^3t_b$, for the values $N=10^6$, 
$r_h/r_b=0.12$ and $M/M_b=6.4\times 10^{-2}$ corresponding to the four 
merged clusters of case 1, and $t_{rh}=2.6\times 10^3t_b$, for the
four merged clusters of case 2, where the only different value is that of
the ratio $r_h/r_b$ which results $0.21$ in this case. In physical units,
this means $t_{rh}=5.8$ Gyr and 
$t_{rh}=13.5$ Gyr in the 2 cases, respectively, assuming $r_b=200$ pc and $M_b=10^9$
M$_\odot$. Consequently, the merged system should stay almost stationary for
a time much longer than that of our present simulations and, more relevantly, 
of the order of the age of galaxies, especially when thinking that
the \lq true \rq number of stars in the system 
is from $64$ to $6.4\times 10^3$ times the one used in the simulations
for $M_b$ in the range $10^9$--$10^{11}\msol$, so to scale all the relaxation
times estimated above for a factor in the range $50$--$3,900$.
If other clusters merge to the center, $N$ 
increases, as likely $r_h$ does, increasing furtherly $t_{rh}$. 
In conclusion, a collisional evolution for the nuclear cluster seems 
unlikely,  but larger scale instabilities cannot be ruled out. 
So, the dynamical evolution of the merger remnant deserves a more extended in time
simulation. 

In the, likely, hipothesis that the NC keeps its stability during subsequent
merging events, the central density $\rho_0$ growth shown by Fig.\ref{dens}
can be reasonably scaled with the number $N_m$ of merging clusters, giving
$\rho_0\simeq 575 N_m$ (case 1) and $\rho_0\simeq 3.7 N_m$ (case 2) in units
of the central galactic density, ${\rho_b}_0$. With $M_b=3\times 10^9 \msol$ and
$r_b=200$ pc as typical values of normal elliptical galaxies, one has
${\rho_b}_0 =375 \msol$ pc$^{-3}$; consequently, for the NC to reach the
high density of the order of  $10^7 \msol$ pc$^{-3}$, which is
typical of the environment of massive accreting objects, $N_m\simeq 46$ (case 1)
and $N_m\simeq 7150$ (case 2) are needed. Even with all the necessary cautions
suggested by the various assumptions done, these numbers (particularly for
case 1) are indicating that a NC formed by merging of individual sub-structures
may actually have reached central densities large enough and in a time short enough
to provide mass for an efficient accretion of a possible massive galactic object,
as suggested in various previous studies \citep[e.g.,][]{capdol93,ct99,cv05}.
In light of this, it is clear the importance of further, more detailed and
complete, study of the actual modes of formation and evolution of a NC via
the merging process (Capuzzo-Dolcetta \& Miocchi, in preparation).


\section{Conclusions}
In this paper we studied the modes of interaction of a sample of few (4) globular clusters 
whose orbital motion is limited to the core of the galactic triaxial field, because
they have experienced a significant and rapid dissipation of the orbital energy by
dynamical friction acting on their large initial masses.
We studied 2 sets of 4 GCs of different initial conditions,
characterised by a higher (case 1) and lower (case 2) central density.
Dynamical friction was properly taken into account by
the triaxial generalization of the classical Chandrasekhar formula \citep{pesce} and
it is the main cause of orbital energy dissipation up to the beginning of the merging
process, when the individual cluster sizes are comparable with the orbital size. 
Since this stage on, the main responsible of the residual orbital energy 
loss is the tidal torque.
The merger is completed in 
$\sim 18$ galactic core crossing times, i.e. in a time much shorter than the Hubble
time and, in any case, small compared to the total orbital decaying time.

The merged system keeps some of the characteristics of the preexisting objects,
attaining a relaxed structure which has, in the case 2, a mildly triaxial shape
(inherited by the
environment) and a halo which is diffusing in the 
external field assuming its phase space properties.
The case 1 results into a merger configuration which conserves 
better the spherical simmetry of the \lq progenitors\rq\ and whose radial density profile
is more concentrated to the center than that expected on the basis of the mere space
superposition of the 4 progenitors.
This is, likely, the consequence of some violent relaxation of the
merged cluster to the state of a quasi-stationary Super Star Cluster,
whose stellar density distribution maintains a King shape and is very similar to
Nuclear Clusters observed at the center of many galaxies. 
Even in the limits of our simulations (both on the number of merging objects
and on the time extension of the integration), we may infer from a comparison
of our results with the characteristics (surface brightness profile, integrated light
and velocity dispersion) of the nuclei of many galaxies \citep{bok02, geha, walch05, cote06}
that such nuclei may have actually formed by the merging 
of few tens (when merging progenitors are quite compact) up to few
hundreds (when progenitors are looser) massive GCs decayed orbitally into the 
inner galactic regions.

As remarkable result, the radial
profile of the line-of-sight velocity dispersion of the merger remnant shows the same
minimum at the galactic center as actually found by observations
of a sample of Virgo cluster nucleated dEs \citep{geha} and of late-type spirals
\citep{walch05}.
Finally, an important theoretical result is that high densities ($\gtrsim 10^7 \msol$ pc$^{-3}$),
critical to allow a significant accretion onto a massive object, may be reached
by mean of a merging of less than $50$ clusters of the type considered in case 1.
This may have important implications on the role of NCs in supplying the innermost
galactic activity.

\section{Acknowledgements}
The main computational resources employed for this work
were provided by CINECA (http://www.cineca.it) thanks to the 
agreement with INAF (http://inaf.cineca.it) under the \emph{\/Key-Project} 
grant \emph{\/inakp002}.
We thank dr. P. Di Matteo for providing us with her subroutine to generate
multi-mass King models, used for initial conditions, and dr. A. Vicari for 
his  fitting routines of the \schw velocity dispersion tensor, 
useful to calculate dynamical friction on the merging GCs.


\begin{figure}
\plotone{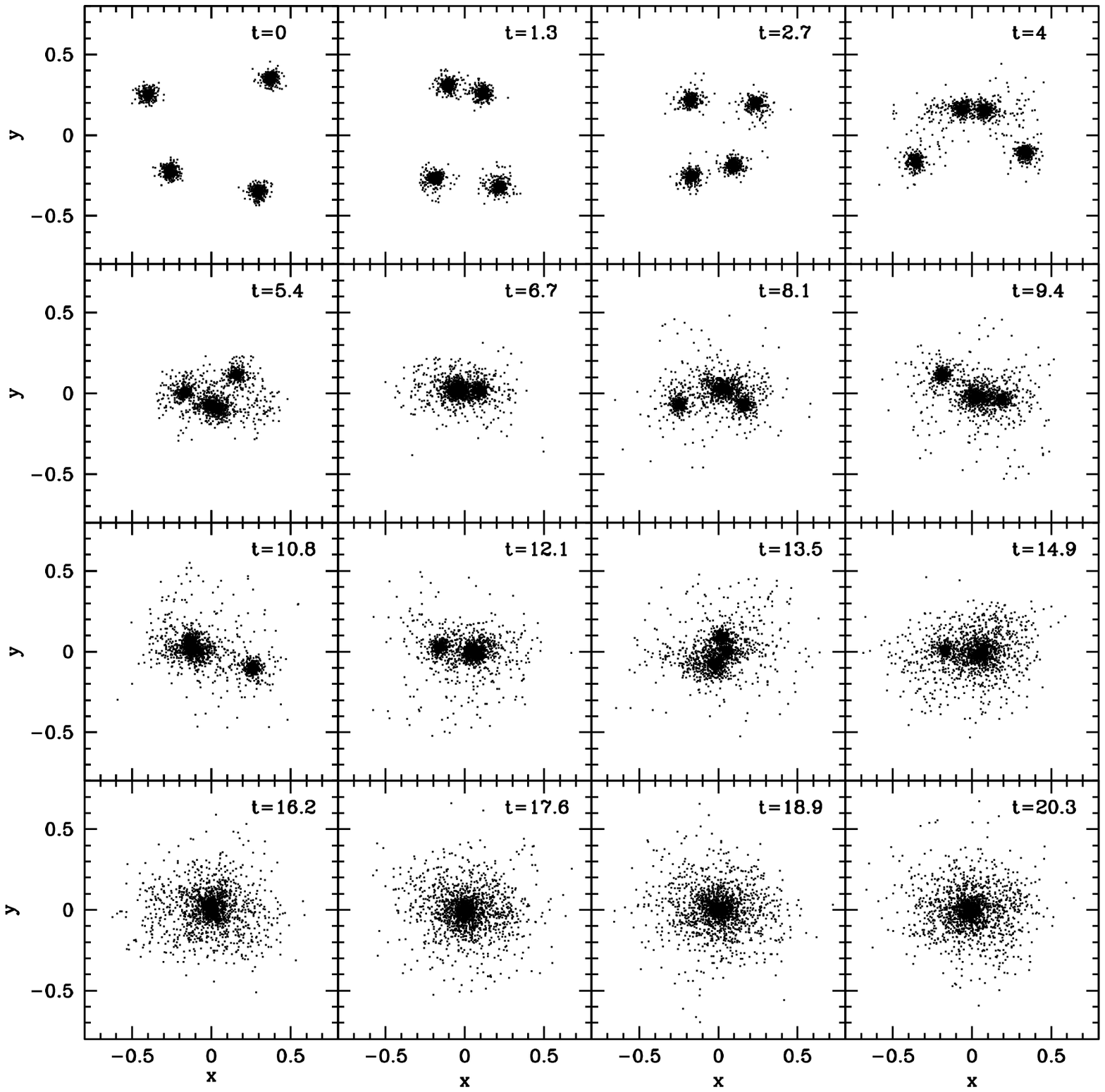}
\caption{Snapshots of the simulation in the case 1
(projection on the $x$--$y$ plane) from $t=0$ to $t=20.3$.
\label{snap1}}
 \end{figure} \clearpage
\begin{figure}
\plotone{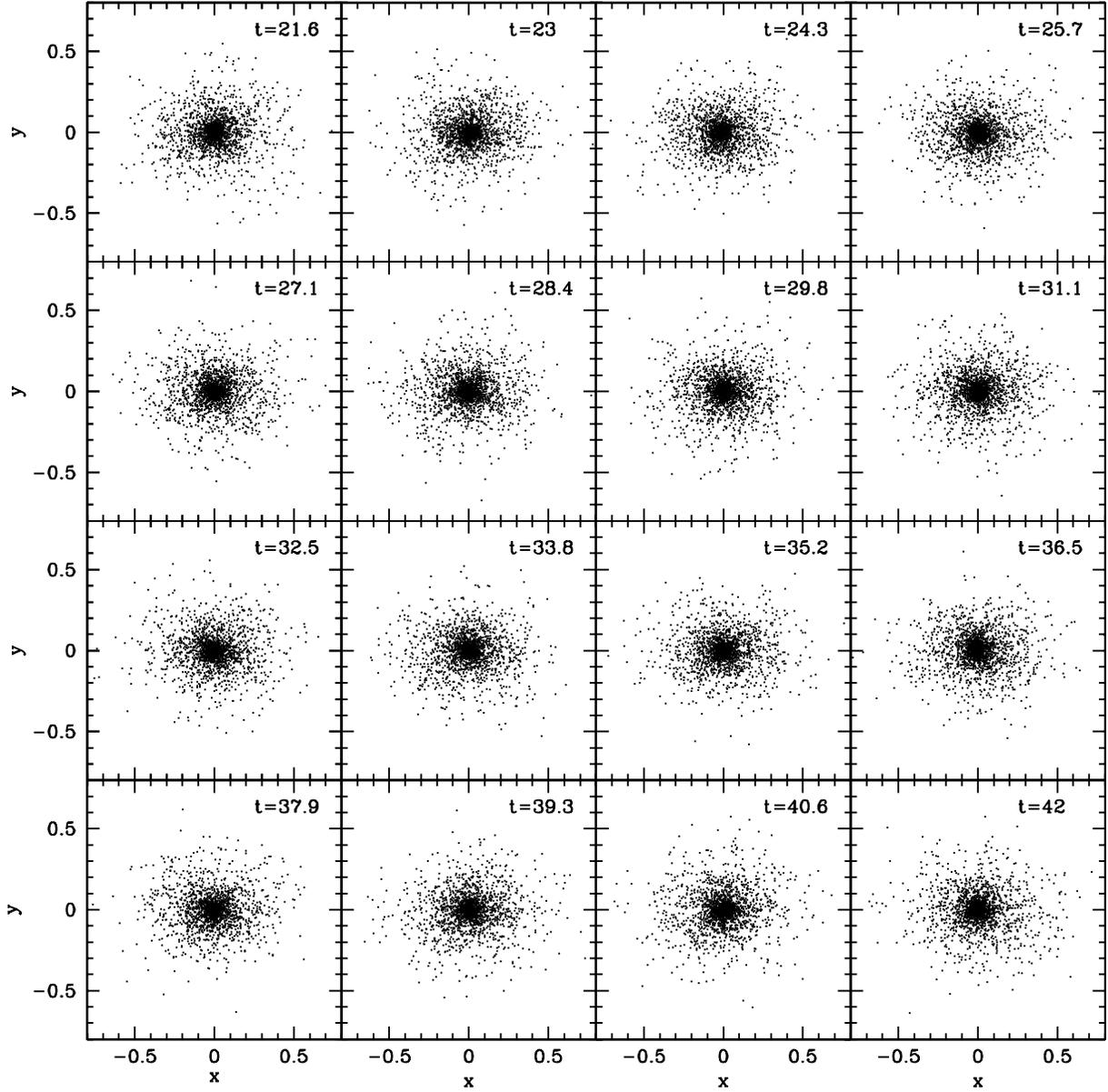}
\caption{Continuation of Fig.~\ref{snap1} from $t=21.6$ to $t=42$.
\label{snap2}}
 \end{figure} \clearpage
\begin{figure}
\plottwo{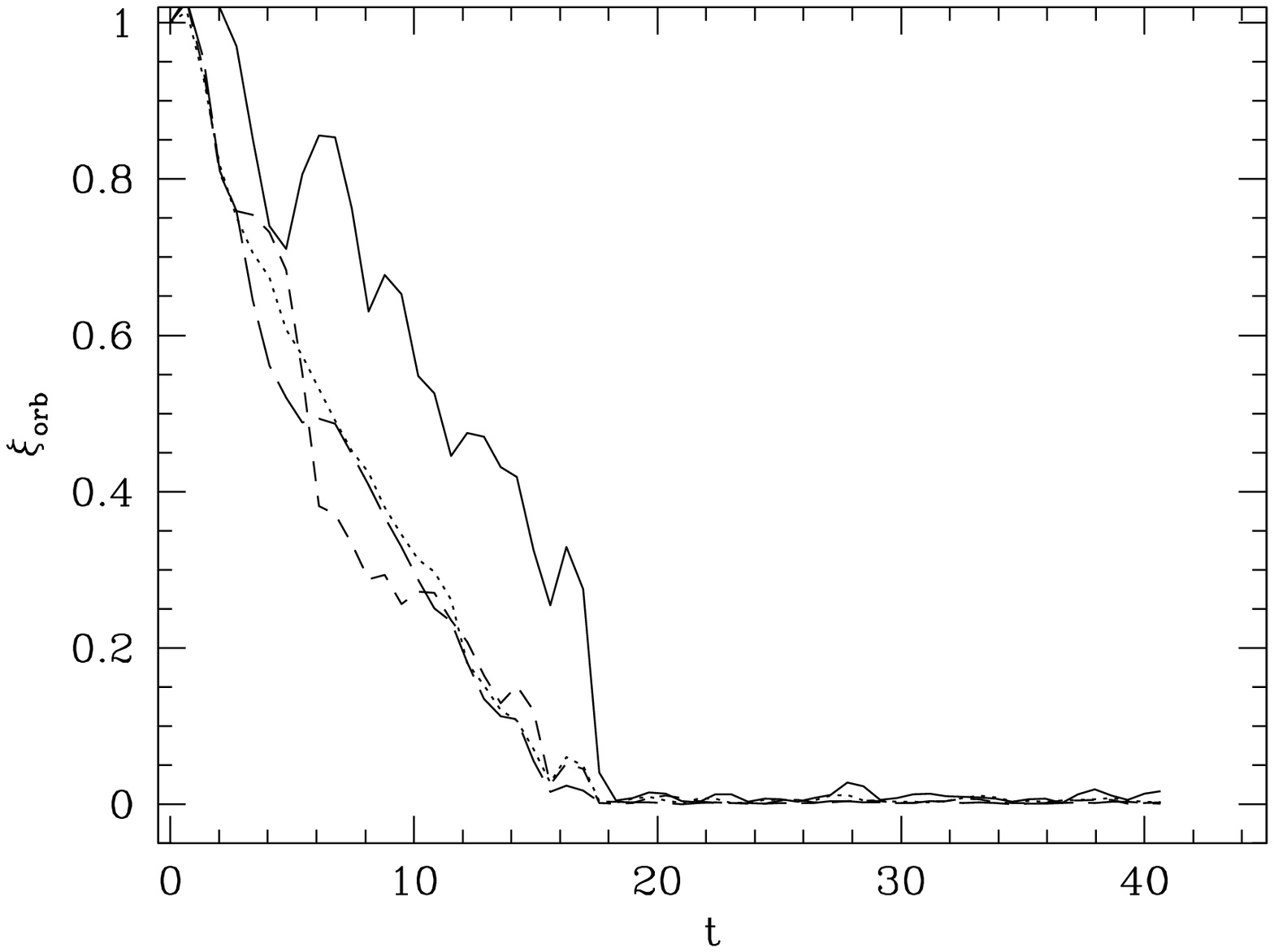}{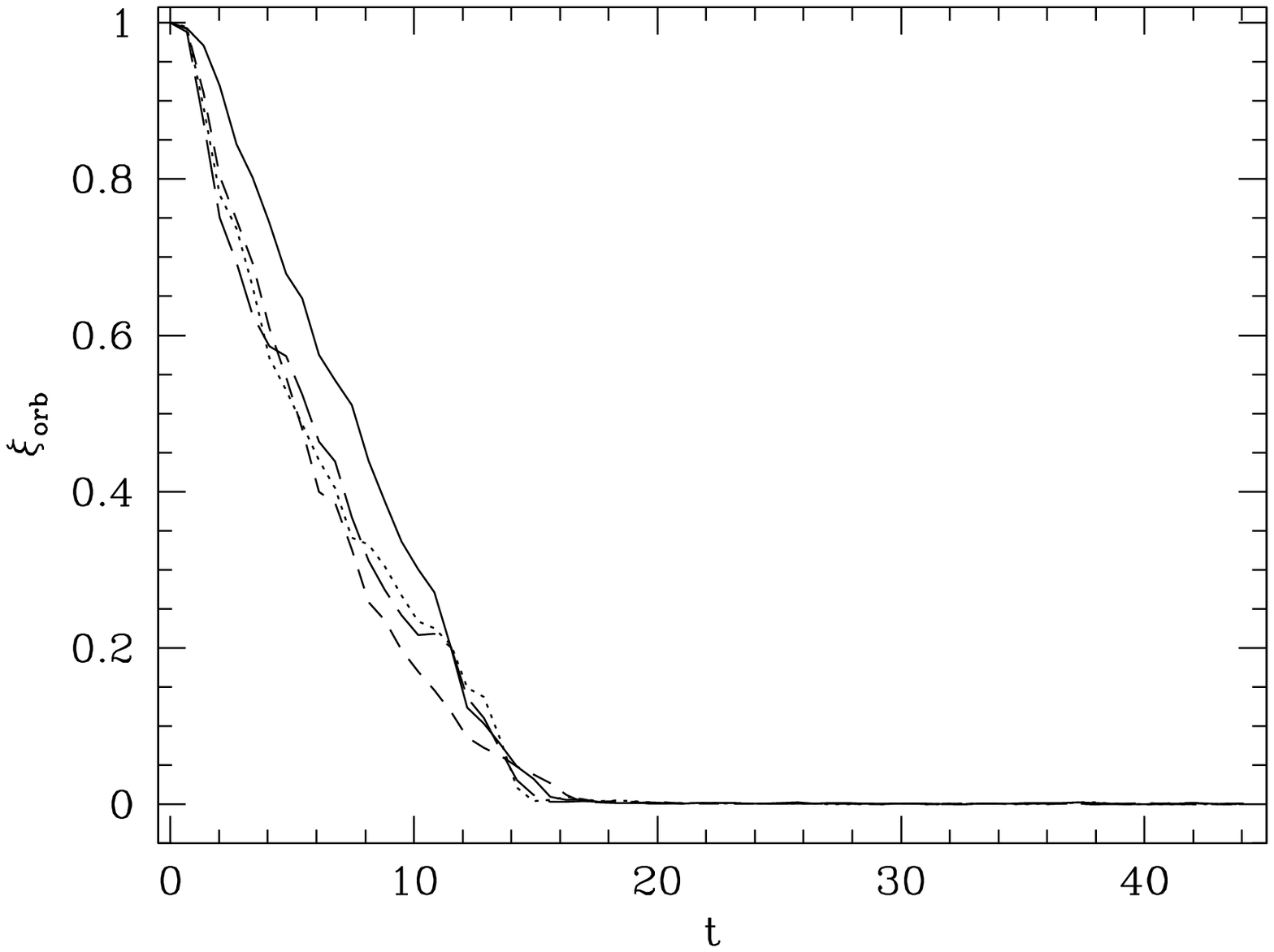}
\caption{Time behaviour of the fraction of the orbital energy kept by the
clusters CD.
Solid line: clusters \a, \al; dotted: \b, \bl; short-dashed: \C, \cl;
long-dashed: \d, \dl.
Left panel: NC$_1$ progenitors; right panel: NC$_2$ progenitors.
\label{decadorb}}
 \end{figure}
\clearpage
\begin{figure}
\plottwo{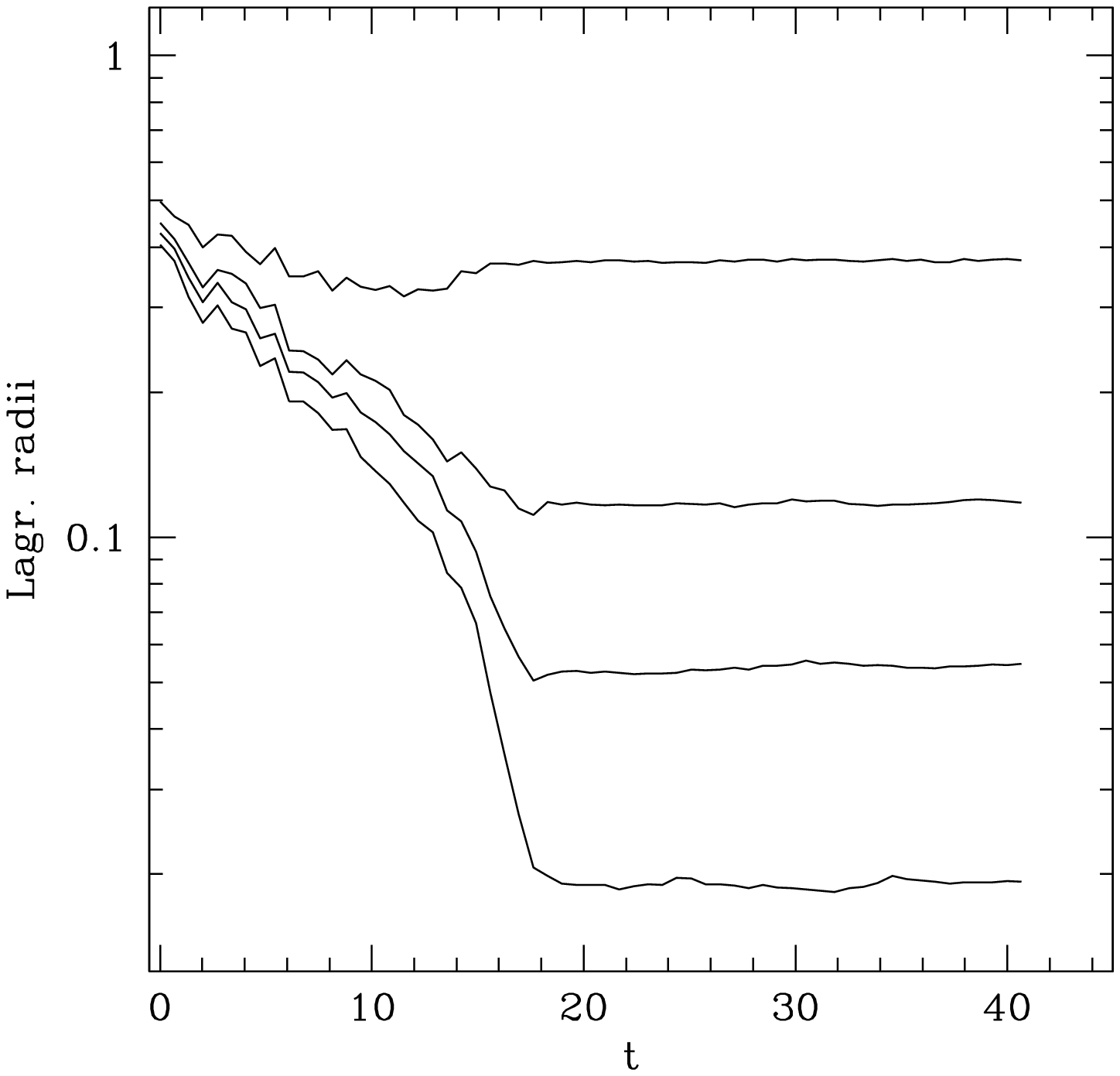}{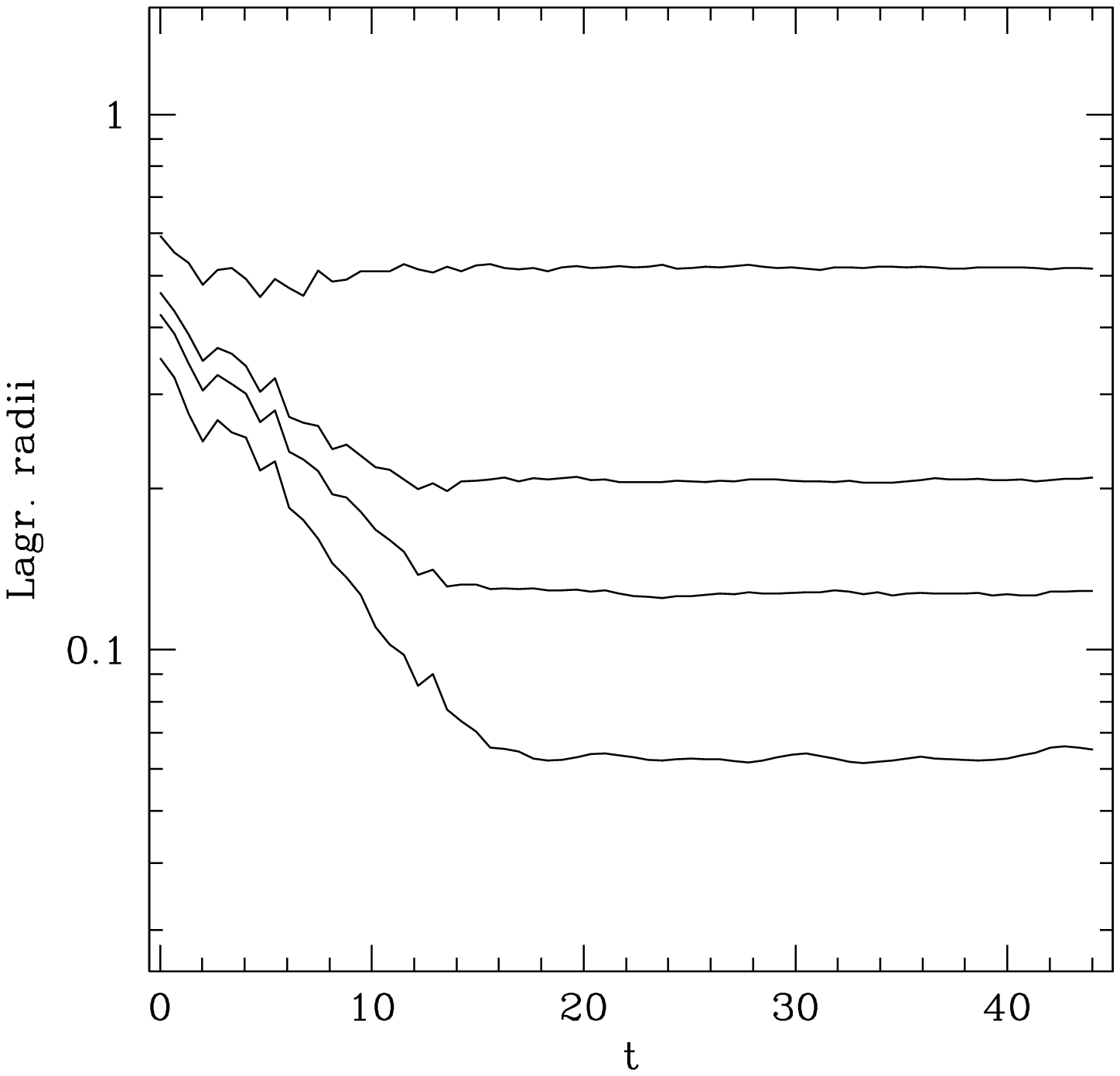}
\caption{Evolution of the Lagrangian radii evaluated with respect to the global CD.
They refer to 10\%, 30\%, 50\% and 90\% of the
total mass of the whole system. Left panel: case 1; right
panel: case 2. \label{rlag}}
 \end{figure} \clearpage
\begin{figure}
\plottwo{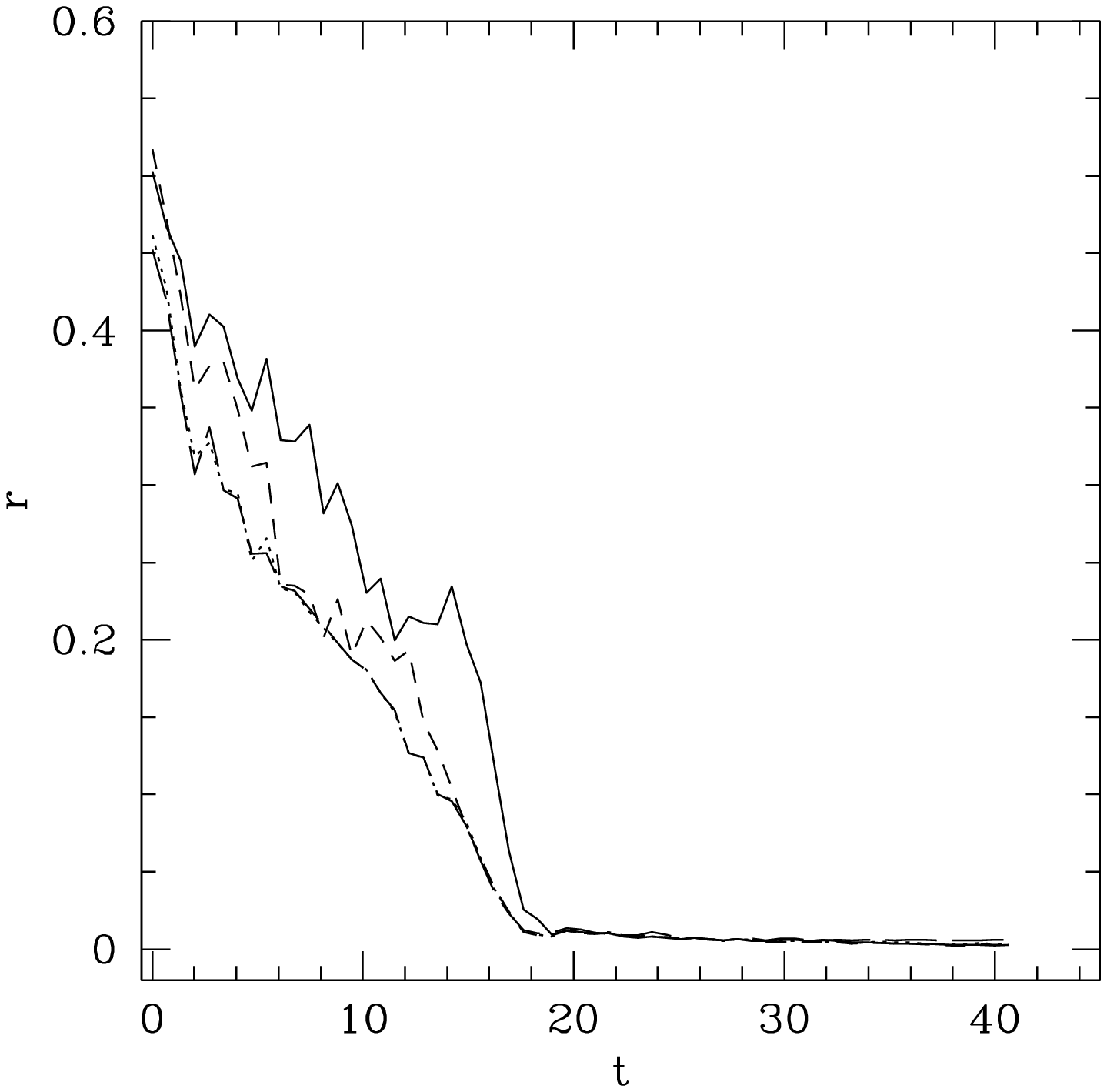}{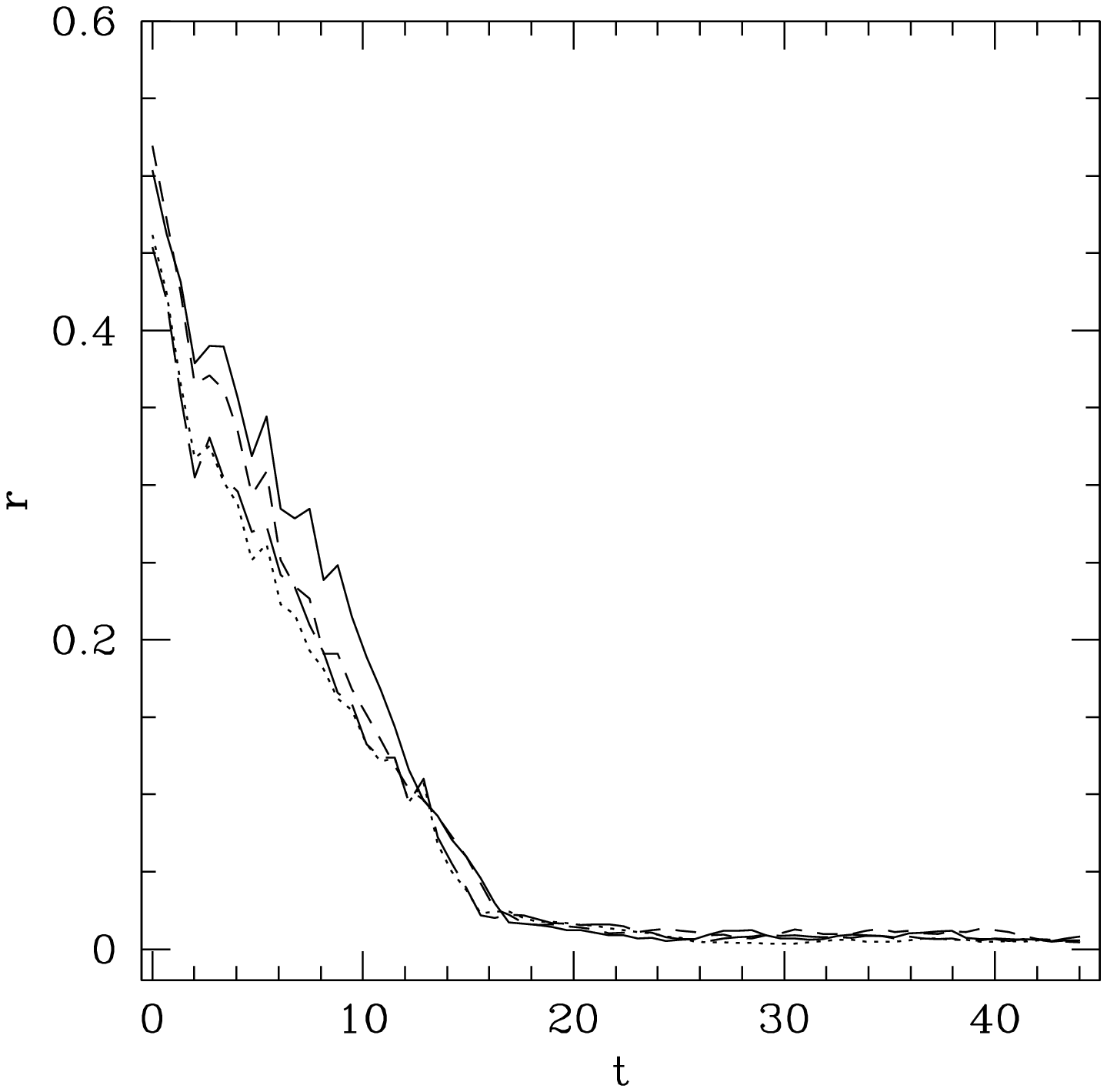}
\caption{Time behaviour of the distance of the clusters CD to the galactic
center.
Left panel: NC$_1$ progenitors; right panel: NC$_2$ progenitors.
Symbols are as in Fig.~\ref{decadorb}\label{distance}}
 \end{figure}
\clearpage
\begin{figure}
\plotone{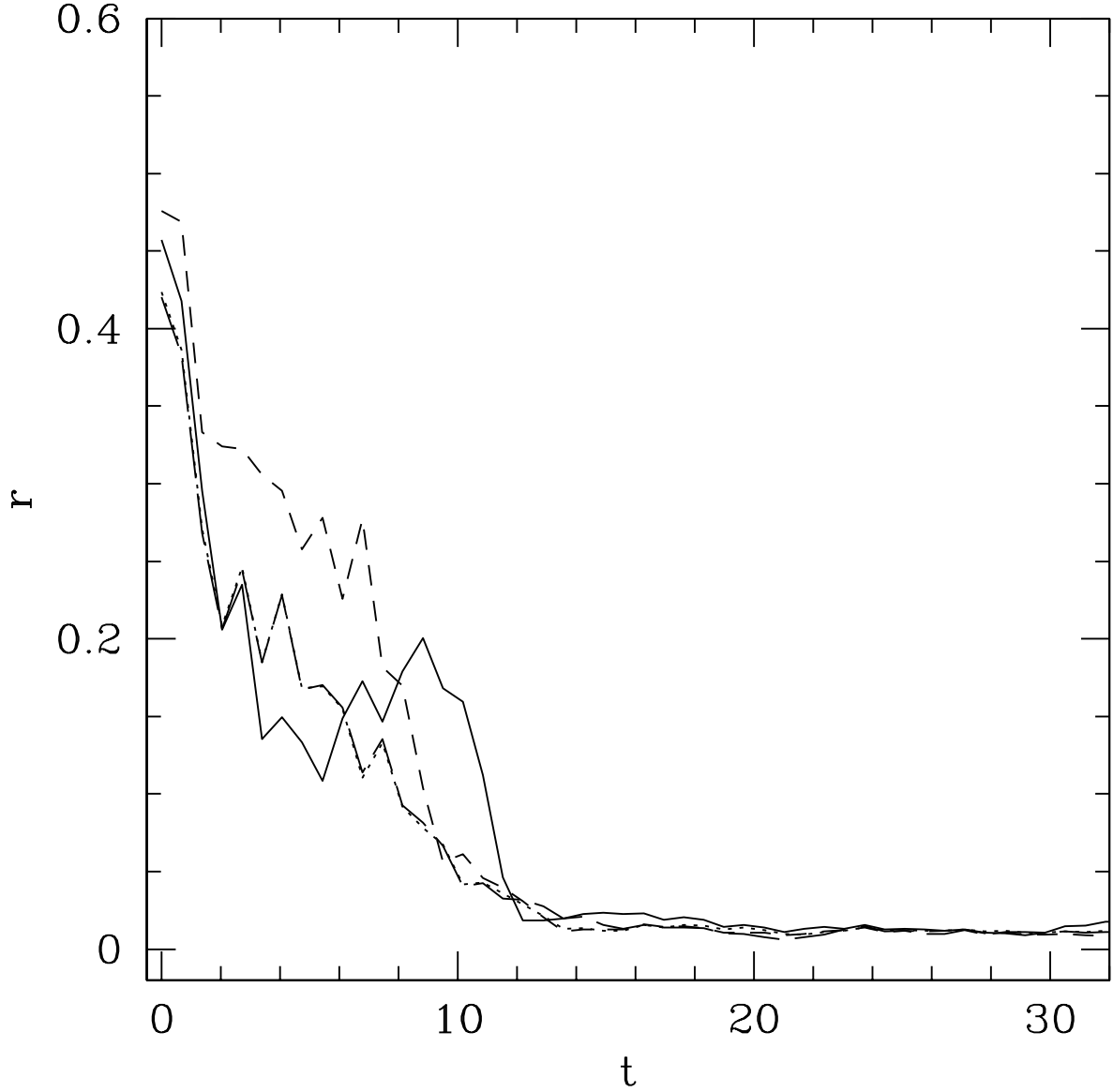}
\caption{As in Fig.~\ref{distance} for the clusters of case 1, but moving
in a particle self-consistent model of the galaxy (from Capuzzo-Dolcetta \&
Miocchi, in preparation).
\label{distance_full}}
 \end{figure}
\clearpage
\begin{figure}
\plotone{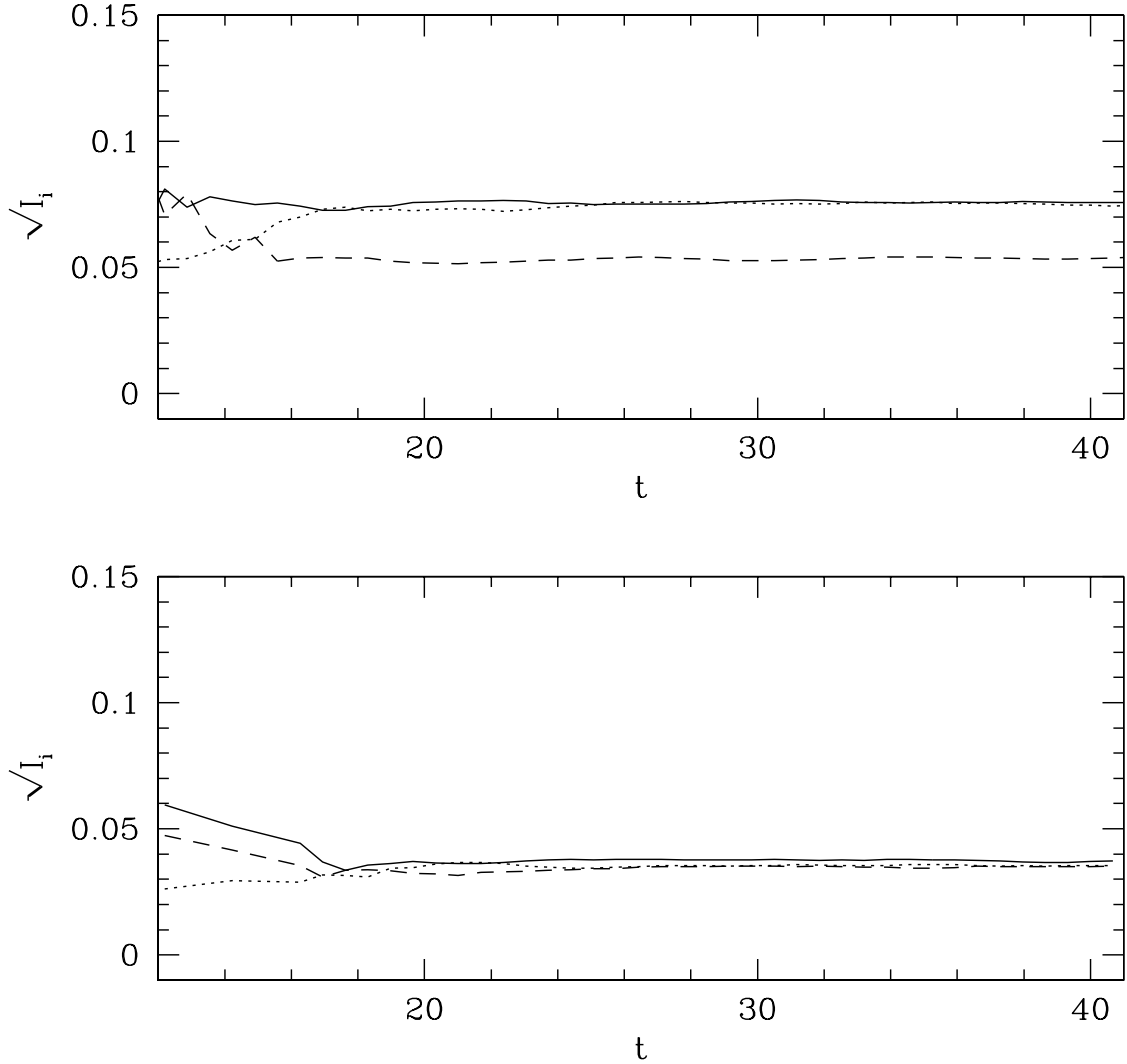}
\caption{Square root of the eigenvalues of the
inertia tensor (solid line: $I_1$, dotted: $I_2$, dashed: $I_3$)
evaluated on the part of the system lying in the sphere centered at the
CD of the whole system and with radius equal to the final NC's
half-mass radius.
Lower panel: case 1; Upper panel: case 2.
\label{eigen}}
\end{figure} \clearpage
\begin{figure}
\plottwo{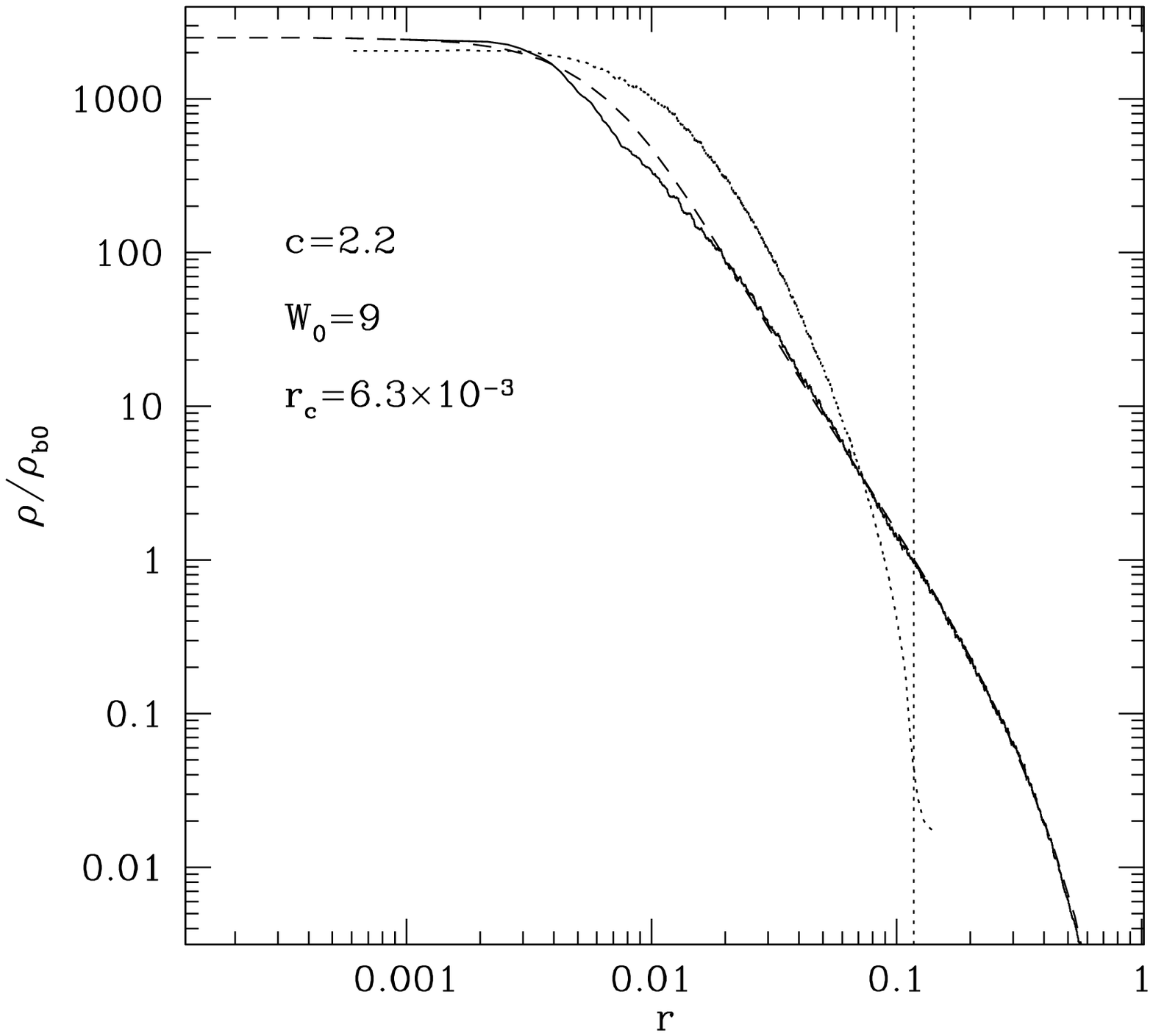}{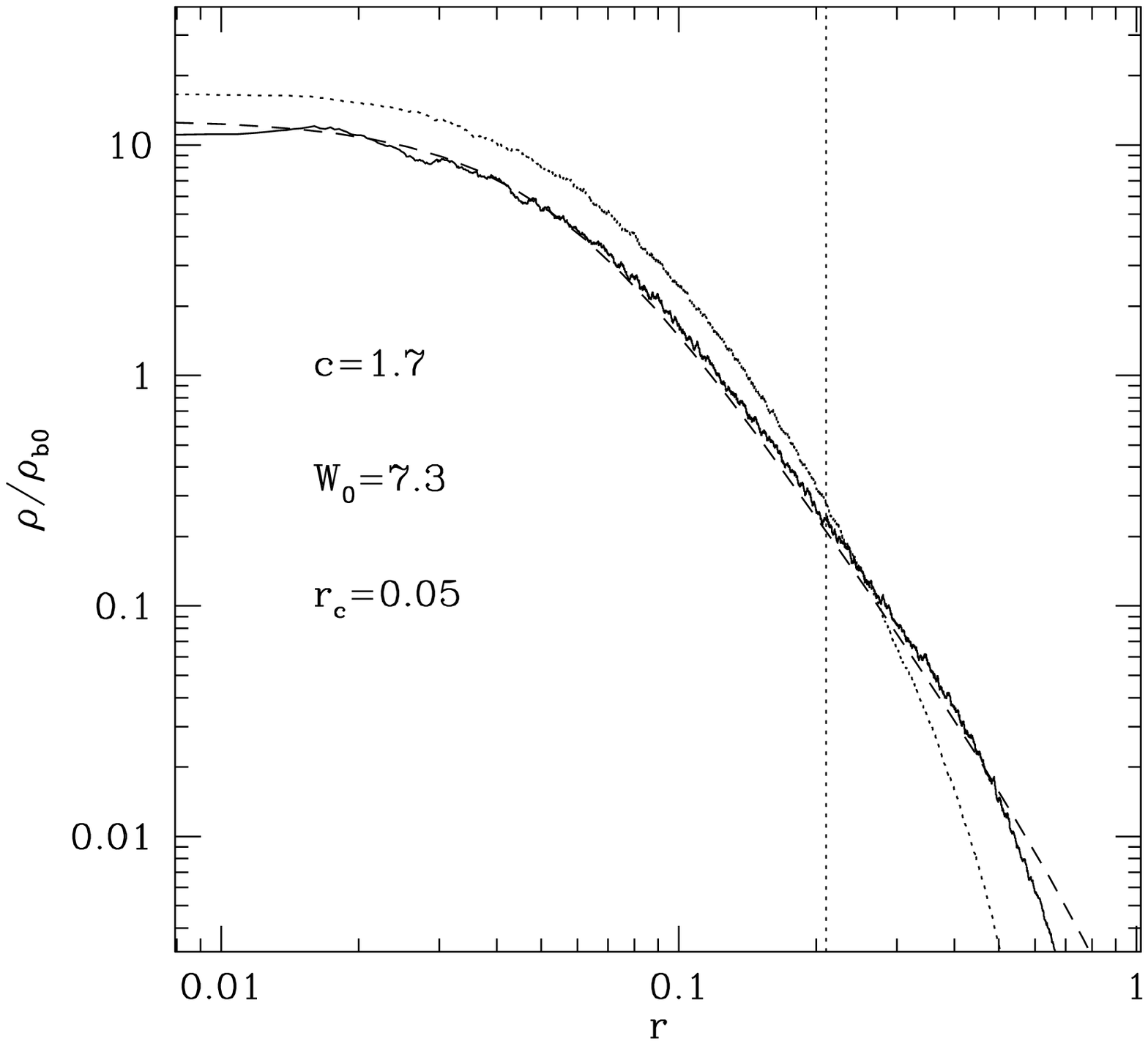}
\caption{Radial volume density profile of the NC at the end of the
simulation (solid line), compared with that corresponding to the sum
of the initial
density distributions of the 4 clusters (dotted line).
Left panel: case 1, right panel: case 2.
The dashed line 
represents the King best-fit to the NC profile.
The structural parameter of the King profile are reported.
The distance $r$ is to the galactic center.  The vertical
dotted line marks the NC half-mass radius. \label{dens}}
\end{figure}
\clearpage
\begin{figure}
\plottwo{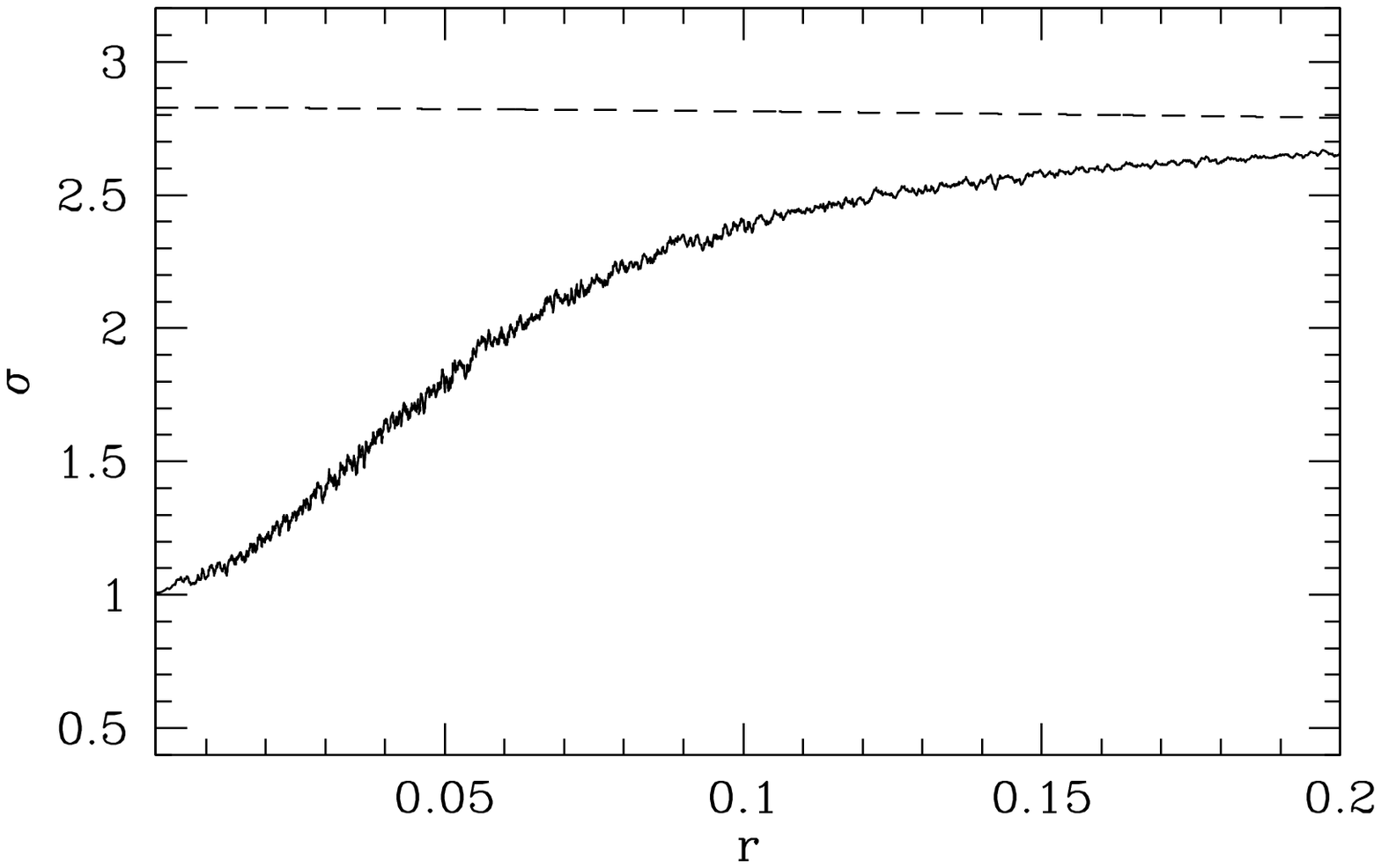}{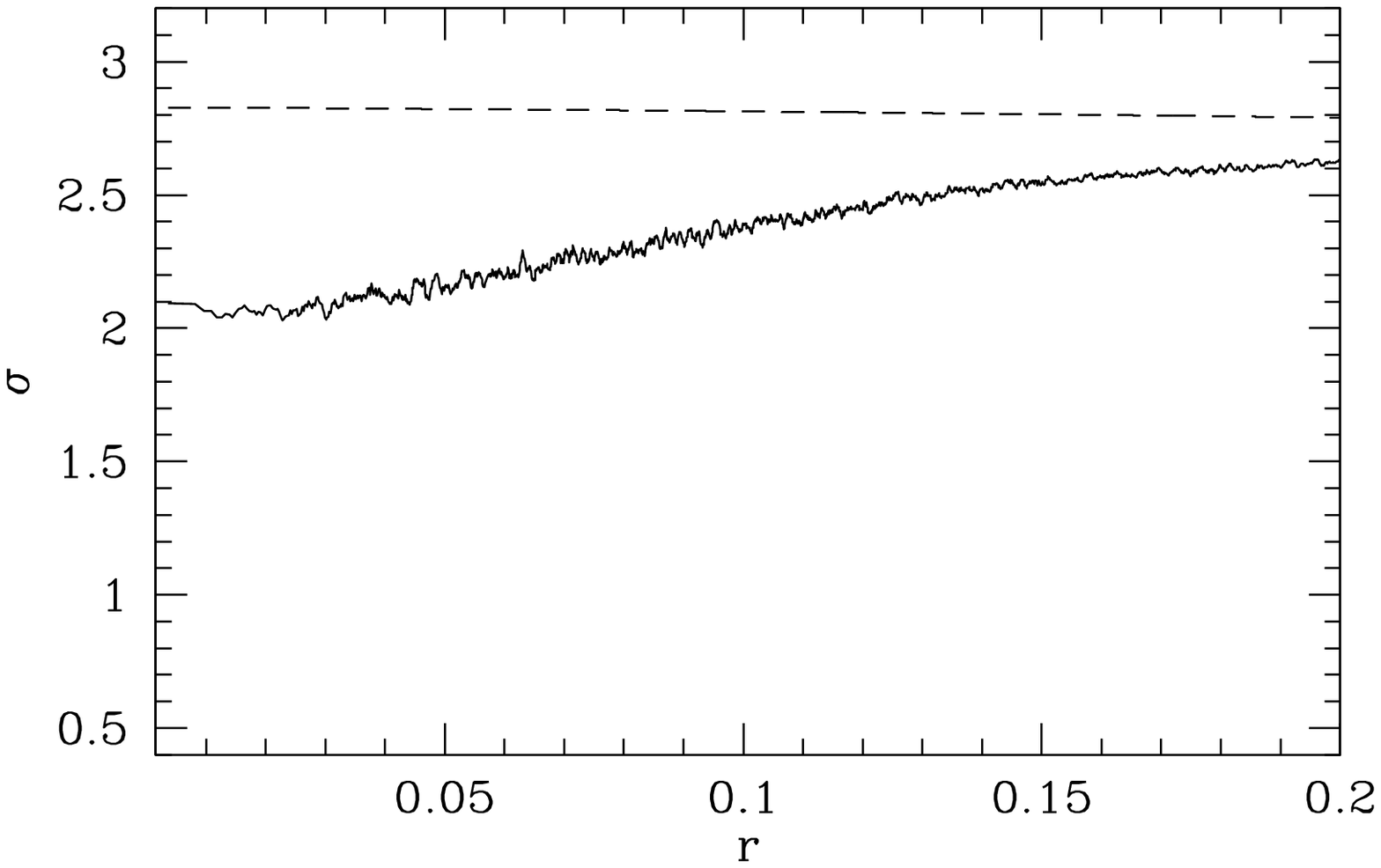}
\caption{Total 3-d velocity dispersion profile.
Left panel: case 1; right panel: case 2.
Dashed curve is the velocity dispersion of the galactic model only.
The total $\sigma$ is computed by summing the galactic model
velocity dispersion plus that of the NC, weighted by the surface density.
\label{veldisp}}
 \end{figure}
\clearpage

\begin{figure}
\plottwo{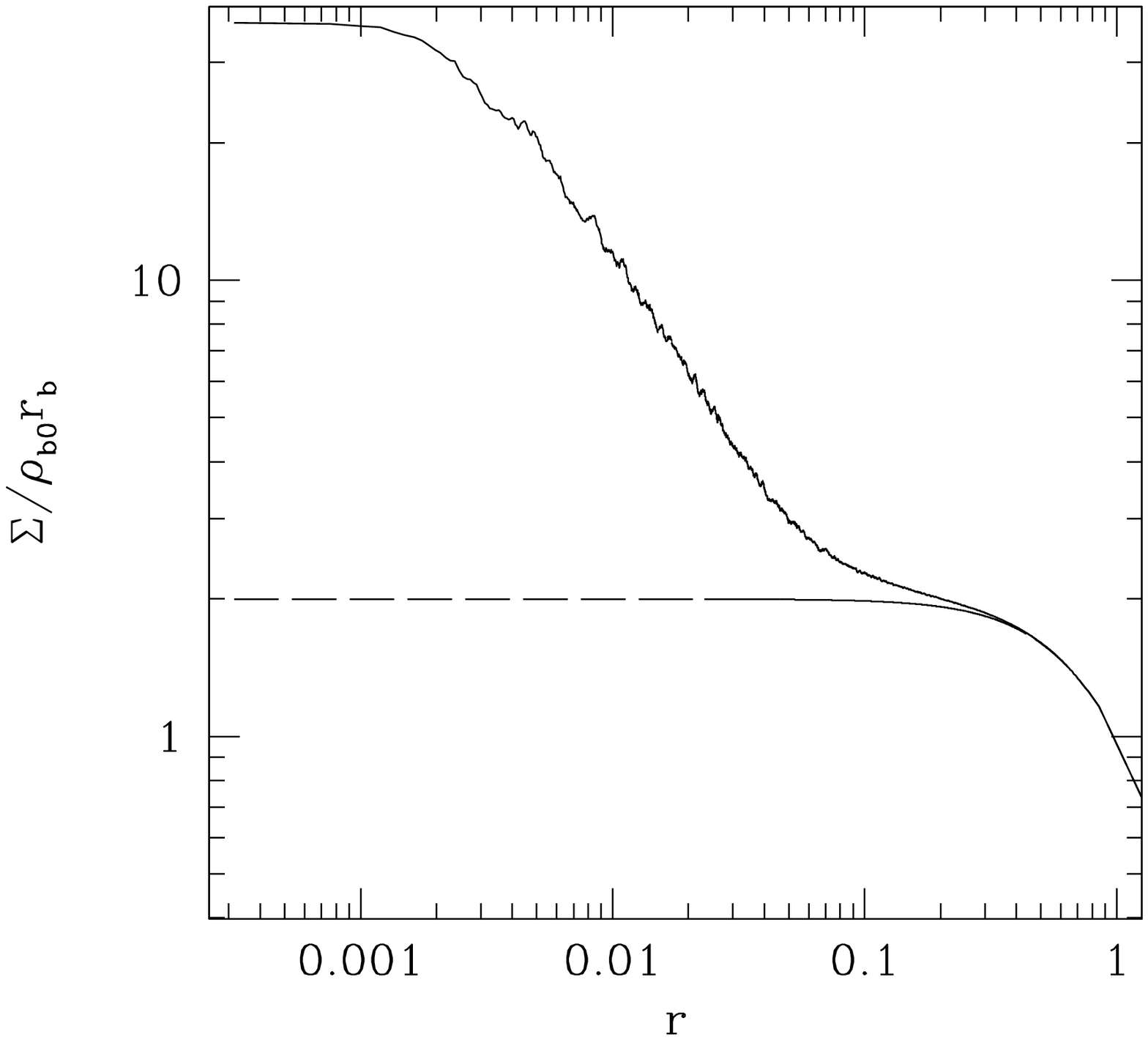}{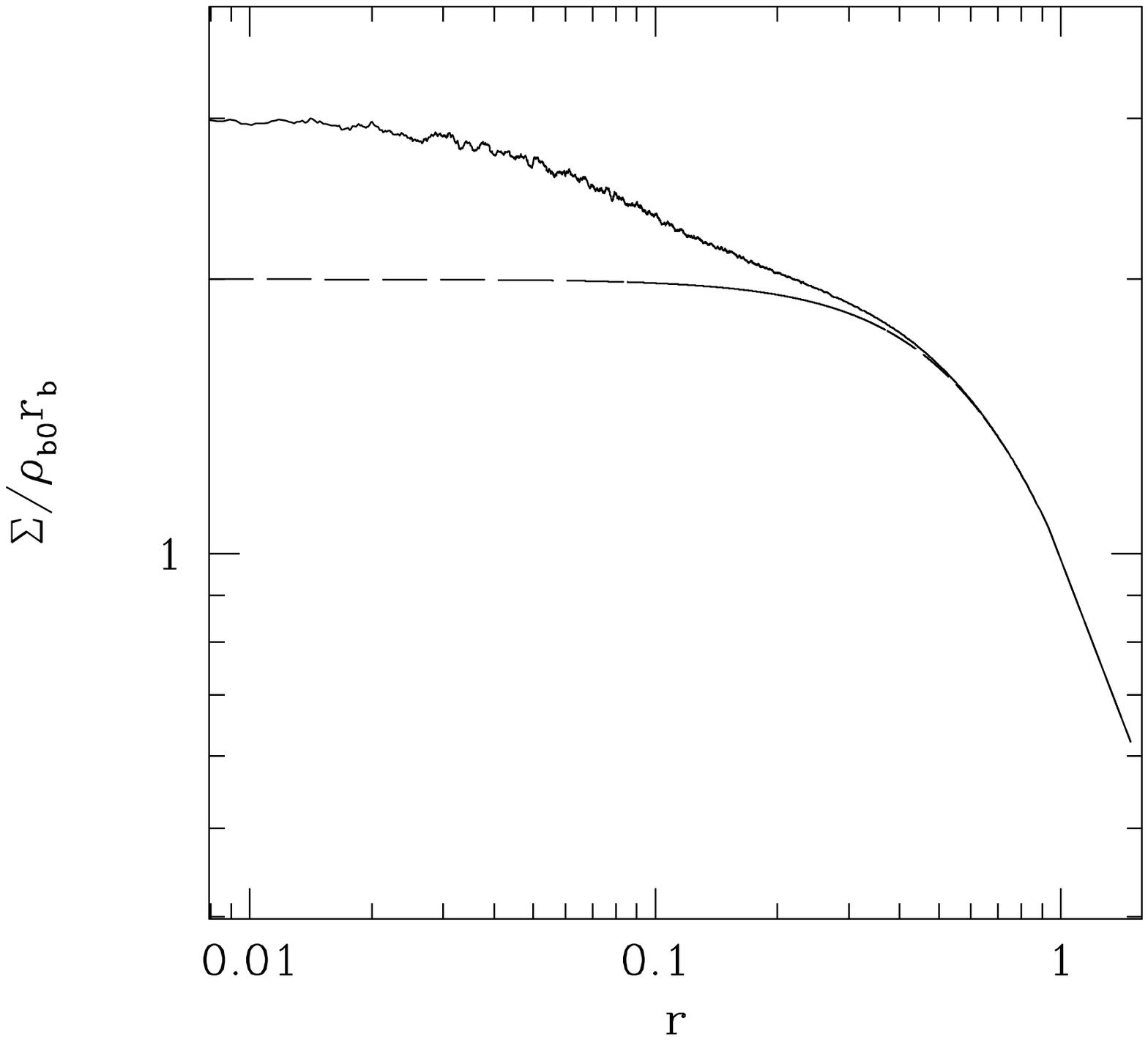}
\caption{ Projected surface density profile for the last NC configuration,
summed to the galaxy background profile (long dashed line) so to give
the total surface density (solid line). Left panel: case 1,
right panel: case 2. \label{dens2}}
 \end{figure}
\clearpage

\begin{deluxetable}{p{0.8cm}ccccccccc}
\tablecaption{Clusters initial parameters.
\label{tab1}}
\tablewidth{0pt}
\tablehead{
\colhead{cluster model} & \colhead{$M$} &\colhead{$r_t$} & \colhead{$c$} &
\colhead{$r_c$} & \colhead{$r_h$} & \colhead{$\rho_0$}
 & \colhead{$t_{ch}$} & \colhead{$\sigma_\mathrm{K}$} & \colhead{$\rho_0/\sigma^3$}
}
\startdata

{\bf \a}&
$1.5\times 10^{-2}$&
$0.16$&
$1.2$&
$1.1\times 10^{-2}$&
$2.3\times 10^{-2}$&
$770$&
$2.6\times 10^{-2}$&
$0.36$&
$3.2\times 10^{3}$\\
{\bf \b}&
$1.7\times 10^{-2}$&
$0.16$&
$1.0$&
$1.5\times 10^{-2}$&
$2.4\times 10^{-2}$&
$410$&
$2.8\times 10^{-2}$&
$0.36$&
$1.7\times 10^{3}$\\
{\bf \C}&
$1.8\times 10^{-2}$&
$0.14$&
$0.99$&
$1.4\times 10^{-2}$&
$2.2\times 10^{-2}$&
$610$&
$2.3\times 10^{-2}$&
$0.41$&
$1.7\times 10^{3}$\\
{\bf \d}&
$1.4\times 10^{-2}$&
$0.14$&
$0.89$&
$1.8\times 10^{-2}$&
$2.4\times 10^{-2}$&
$250$&
$3.3\times 10^{-2}$&
$0.34$&
$1.2\times 10^{3}$\\
{\bf \al}&
$1.5\times 10^{-2}$&
$0.77$&
$1.2$&
$5.3\times 10^{-2}$&
$0.11$&
$6.1$&
$0.30$&
$0.16$&
$290$\\
{\bf \bl}&
$1.7\times 10^{-2}$&
$0.78$&
$1.0$&
$7.3\times 10^{-2}$&
$0.12$&
$3.3$&
$0.31$&
$0.16$&
$150$\\
{\bf \cl}&
$1.8\times 10^{-2}$&
$0.68$&
$0.99$&
$6.9\times 10^{-2}$&
$0.11$&
$4.9$&
$0.26$&
$0.18$&
$160$\\
{\bf \dl}&
$1.4\times 10^{-2}$&
$0.72$&
$0.89$&
$9.3\times 10^{-2}$&
$0.12$&
$2.0$&
$0.37$&
$0.15$&
$110$\\
\enddata

\tablecomments{Parameters list for the initial cluster models,
expressed in galactic units.
Reported are: the GC mass ($M$), the limiting radius ($r_t$), the King
concentration coefficient ($c$), the King radius ($r_c$), the half-mass radius ($r_h$),
the central density ($\rho_0$), the half-mass crossing time [$t_{ch}\equiv [r_h^3/(GM)]^{1/2}$],
the King velocity parameter ($\sigma_\mathrm{K}$), and the central phase-space density estimate (where
$\sigma=\sqrt{3}\sigma_\mathrm{K}$).
}
\end{deluxetable}

\clearpage

\begin{deluxetable}{p{0.8cm}cccccccc}
\tablecaption{NCs parameters.
\label{tab2}}
\tablewidth{0pt}
\tablehead{
\colhead{Nuclear cluster} & \colhead{$M$} & \colhead{$c$} &
\colhead{$r_c$} & \colhead{$r_h$} & \colhead{$\rho_0$}
 & \colhead{$t_{ch}$} & \colhead{$\sigma_0$} & \colhead{$\rho_0/\sigma_0^3$}
}
\startdata

{\bf NC$_1$}&
$6.4\times 10^{-2}$&
$2.2$&
$6.3\times 10^{-3}$&
$0.12$&
$2.3\times 10^{3}$&
$0.16$&
$0.90$&
$3.1\times 10^{3}$\\  
{\bf NC$_2$}&
$6.4\times 10^{-2}$&
$1.7$&
$5.0\times 10^{-2}$&
$0.21$&
$15$&
$0.38$&
$0.50$&
$120$\\ 
\enddata

\tablecomments{Parameters list for the NCs in the last configuration, expressed
in galactic units.
Reported are: the total mass ($M$), the
concentration coefficient ($c$) and the King radius ($r_c$) of the best-fit King model, the
half-mass radius ($r_h$), the central density ($\rho_0$), the half-mass crossing time
($t_{ch}$),
the central velocity dispersion ($\sigma_0$) 
and the central phase-space density estimate.
}
\end{deluxetable}

\clearpage

\begin{deluxetable}{p{2.cm}rrr}
\tablecaption{Orbital initial conditions for the clusters.
\label{initialorbits}}
\tablewidth{0pt}
\tablehead{
\colhead{cluster model} & \colhead{$x_0$} & \colhead{$y_0$} & \colhead{$z_0$}
}


\startdata

{\bf \a} -- {\bf \al}&
$-0.4$&
$0.25$&
$-0.35$\\

{\bf \b} -- {\bf \bl}&
$0.3$&
$-0.35$&
$0.31$\\

{\bf \C} -- {\bf \cl}&
$0.375$&
$0.35$&
$-0.325$\\

{\bf \d} -- {\bf \dl}&
$-0.26$&
$-0.225$&
$0.425$\\
\enddata

\tablecomments{Initial conditions of the orbits of the 4 clusters.
All started with zero initial velocity.}

\end{deluxetable}


\begin{thebibliography}{}
\bibitem[Ashman \& Zepf, 1998]{ashzepf98}Ashman, K. M.,  Zepf, S. E. 1998,
Globular cluster systems New York: Cambridge University Press, 1998. (Cambridge
astrophysics series ; 30) 
\bibitem[Babul \& Rees, 1992]{babrees92}Babul, A., \& Rees, M.J. 1992, \mnras, 255, 346
\bibitem[Baumgardt et al., 2003]{baumg03}Baumgardt, H., Makino, J., Hut,
P., McMillan, S., Portegies Zwart, S. 2003, \apj, 589L, 25	
\bibitem[Bekki et al., 2004]{bekki04}  Bekki, K., Couch, W.J., Drinkwater, M.J.
\& Shioya, Y. 2004, ApJ 610, L13
\bibitem[Bertola et al., 1991]{bertola91}Bertola, F., Vietri, M, \& Zeilinger, W.W. 1991, \apj, 374, L13
\bibitem[Binney \& Tremaine, 1987]{bib4} Binney, J., \& Tremaine, S. 1987,
Galactic Dynamics, Princeton Univ. Press
\bibitem[B\"oker et al., 2002]{bok02}
B\"oker, T., Laine, S., van der Marel, R. P., Sarzi, M., Rix, H.-W., Ho, L. C.,
 Shields, J. C. 2002, \aj, 123, 1389
\bibitem[B\"oker et al., 2004]{bok04}
B\"oker, T., Sarzi, M., McLaughlin,
D. E., van der Marel, R. P., Rix, H.-W., Ho, L. C., Shields,
J. C. 2004, \aj, 127, 105	
\bibitem[Capuzzo-Dolcetta, 1993]{capdol93}Capuzzo-Dolcetta, R. 1993, \apj, 415, 616
\bibitem[Capuzzo-Dolcetta et al., 2005]{nostro} Capuzzo-Dolcetta, R., Di Matteo,
P., \& Miocchi, P. 2005, AJ, 129, 1906
\bibitem[Capuzzo-Dolcetta \& Tesseri, 1997]{ct97}Capuzzo-Dolcetta, R., \&
Tesseri, A. 1997, \mnras, 292, 808
\bibitem[Capuzzo-Dolcetta \& Tesseri, 1999]{ct99}Capuzzo-Dolcetta, R., \&
Tesseri, A. 1999, \mnras, 308, 961
\bibitem[Capuzzo-Dolcetta \& Vicari, 2005]{cv05}Capuzzo-Dolcetta, R., \& Vicari,
A. 2005, \mnras, 356, 899
\bibitem[Casertano \& Hut, 1985]{cashut}Casertano, S., \& Hut, P. 1985, \apj,
298, 80
\bibitem[Chandrasekhar, 1943]{chandra}Chandrasekhar, S. 1943, \apj, 97, 255
\bibitem[C\^ot\`e et al., 2006]{cote06}C\^ot\`e, P. et al. 2006, \apjs, 165, 57
\bibitem[Davies et al., 2001]{davies01}Davies, R. L. , Kuntschner, H. ,
Emsellem, E., Bacon, R., Bureau, M., Carollo, C.M., Copin, Y., Miller, B.W., Monnet, G.,
Peletier, R.F., Verolme, E.K., \& de Zeeuw, P.T. 2001, \apj, 548, L33
\bibitem[de Zeeuw \& Merritt, 1983]{zeeuw}de Zeeuw, T., \& Merritt, D. 1983,
\apj, 267, 571
\bibitem[Fellhauer \& Kroupa, 2002]{fell02}Fellhauer, M., \& Kroupa, P. 2002,
\mnras, 330, 642
\bibitem[Fellhauer \& Kroupa, 2005]{fell05}Fellhauer, M., \& Kroupa, P. 2005,
\mnras, 359, 223
\bibitem[Fellhauer \& Lin, 2007]{fell07}Fellhauer, M., \& Lin, D.N.C. 2007,
\mnras, 375, 604
\bibitem[Fujii et al., 2007]{funato}Fujii, M., Iwasawa, M., Funato, Y., \& Makino, J.
2007, submitted to \apj (astro-ph/0708.3719)
\bibitem[Geha et al., 2002]{geha} Geha, M., Guhathakurta, P., \& R. P. van der Marel, R.P. 2002,
\aj, 124, 3073
\bibitem[Genzel et al., 2003]{genz03} Genzel, R., et al. 2003, \apj, 594, 812
\bibitem[Graham \& Guzm\'an, 2003]{guzman} Graham, A. W., \& Guzm\'an, R. 2003, \aj,
125, 2936
\bibitem[Harris, 1986]{harris86}Harris, W.E. 1986, \aj, 91, 822
\bibitem[Harris, 1991]{harris91}Harris, W.E. 1991, \araa, 29, 543
\bibitem[Harris, 2001]{harris01}Harris, W.E. 2001, in Labhardt L., Binggeli B.
eds., Saas-Fee Advanced Course 28, Star Clusters. Springer-Verlag, Berlin, p.223
\bibitem[Ha\c segan et al., 2005]{hasegan}
Ha\c segan, M., et al. 2005, \apj, 627, 203
\bibitem[Kissler-Patig et al., 2006]{kissler}Kissler-Patig, M., Jord\'an, A., \&
Bastian, N. 2006  \aap, 448, 1031
\bibitem[King, 1966]{king66}King, I.R. 1966, \aj, 71, 64
\bibitem[Matthews \& de Grijs, 2004]{matthews04}Matthews, L.D., \& de Grijs, R.
2004, \aj, 128, 137
\bibitem[McCrady, 2004]{McCr}McCrady, N. 2004, AAS~Meeting, 205, 85
\bibitem[Mihos \& Hernquist, 1994]{mihhernq94} Mihos, J.C., \& Hernquist, L.
1994, ApJ, 437, L47
\bibitem[Miocchi \& Capuzzo-Dolcetta, 2002]{bib2} Miocchi, P., \&
Capuzzo-Dolcetta, R. 2002, \aap, 382, 758
\bibitem[Miocchi et al., 2006]{miocchi05} Miocchi, P., Capuzzo-Dolcetta, R., Di
Matteo, P., \& Vicari, A. 2006, \apj, 644, 940 (Paper I)
\bibitem[Oh \& Lin, 2000]{oh00}Oh, K.S., \& Lin, D.N.C. 2000, \apj, 543, 620
\bibitem[\protect\citeauthoryear{Pesce, Capuzzo-Dolcetta \& Vietri}{Pesce et al.}{1992}]
{pesce}Pesce, E., Capuzzo-Dolcetta, R., \& Vietri,
M. 1992, \mnras, 254, 466
\bibitem[Rossa et al., 2006]{rossa06}
Rossa, J., van der Marel, R. P., Boeker, T., Gerssen, J., Ho, L. C.,
Rix, H.-W., Shields, J. C., \& Walcher, C. J. 2006, \aj, 132, 1074
\bibitem[Schwarzschild, 1979]{schw}Schwarzschild, M. 1979, \apj, 232, 236
\bibitem[Silk et al., 1987]{silkws87}Silk, J., Wyse, R.F.G., \& Shields, G.A.
1987, ApJ, 332, L59
\bibitem[Spitzer, 1987]{spitzer} Spitzer, L. Jr. 1987, Dynamical Evolution of
Globular Clusters (Princeton: Princeton University Press)
\bibitem[Statler et al., 2004]{statler04} Statler, T.S., Emsellem, E., Peletier,
R. F., \& Bacon, R.
 2004, \mnras, 353, 1
\bibitem[Tremaine et al., 1975]{tremetal75}Tremaine, S., Ostriker, J.P., \&
Spitzer, L. Jr. 1975, \apj, 196, 407
\bibitem[van den Bergh, 1986]{vdb86}van den Bergh, S. 1986, AJ, 91, 271
\bibitem[van der Marel et al., 2007]{vandermarel}
 van der Marel, R. P., Rossa, J., Walcher, C. J., Boeker, T., Ho, L. C.,
 Rix, H.-W., \& Shields, J. C. 2007, Proceedings of IAU Symp. 241,
''Stellar Populations as Building Blocks of Galaxies``, A. Vazdekis and R.
Peletier., eds. (astro-ph/0702433)
\bibitem[Wagner, 1988]{wagner88}Wagner, S.J., Bender, R., \& Moellenhoff, C.
1988, \aap, 195, L5
\bibitem[Walcher et al., 2005]{walch05} Walcher, C. J., van der Marel, R. P.,
McLaughlin, D., Rix, H.-W., B\"oker, T., H\"aring, N., Ho, L.C., Sarzi, M.,
\& Shields, J. C. 2005, \apj, 618, 237
\bibitem[Walcher et al., 2006]{walch06} Walcher, C. J.,
B\"oker, T., Charlot, S., Ho, L. C., Rix, H.-W., Rossa, J., Shields, J. C.,
\& van der Marel, R. P. 2006, \apj, 649, 692
\bibitem[Wehner \& Harris, 2006]{wehner}
Wehner, E.H, \& Harris, W.E. 2006, \apj, 644, L17
\bibitem[White, 1978]{white78}
White, S.D.M. 1978, \mnras, 184, 185
\end{thebibliography}
\end{document}